\documentclass[aps,superscriptaddress,onecolumn,pre,longbibliography,floatfix,notitlepage]{revtex4-1}
\usepackage[unicode=true,pdfusetitle, bookmarks=true,bookmarksnumbered=false,bookmarksopen=false, breaklinks=false,pdfborder={0 0 0},backref=false,colorlinks=false] {hyperref}
\hypersetup{colorlinks,linkcolor=myurlcolor,citecolor=myurlcolor,urlcolor=myurlcolor}
\usepackage{braket,colortbl,amsthm,amsmath,amssymb,txfonts,graphicx}
\definecolor{myurlcolor}{rgb}{0,0,0.7}
\usepackage{cleveref}
\usepackage{soul}
\usepackage{subfigure}
\newcommand{\ketbra}[2]{|{#1} \rangle \langle {#2} |}
\theoremstyle{plain}
\usepackage{indentfirst}

\newcommand{\bb}{\textcolor{blue}}

\newcommand{\bblk}{\textcolor{black}}

\begin{document}
\title{Sensing multiatom networks in cavities via photon-induced excitation resonance}

\author{Pritam Chattopadhyay\footnote{Corresponding author}}
\thanks{contributed equally}
\email{pritam.chattopadhyay@weizmann.ac.il}
\affiliation{Department of Chemical and Biological Physics \& AMOS,
Weizmann Institute of Science, Rehovot 7610001, Israel}
\author{Avijit Misra}
\thanks{contributed equally}
\email{avijitmisra@iitism.ac.in}
\affiliation{Department of Chemical and Biological Physics \& AMOS,
Weizmann Institute of Science, Rehovot 7610001, Israel}
 \affiliation{Department of Physics, Indian Institute of Technology (ISM), Dhanbad, Jharkhand 826004, India}
 
\author{Saikat Sur}
\thanks{contributed equally}
\email{saikat.sur@weizmann.ac.il}
\affiliation{Department of Chemical and Biological Physics \& AMOS,
Weizmann Institute of Science, Rehovot 7610001, Israel}

\author{David Petrosyan}
\email{dap@iesl.forth.gr}
\affiliation{Institute of Electronic Structure and Laser, FORTH, GR-70013 Heraklion, Greece}

\author{Gershon Kurizki}
\email{gershon.kurizki@weizmann.ac.il}
\affiliation{Department of Chemical and Biological Physics \& AMOS,
Weizmann Institute of Science, Rehovot 7610001, Israel}

\begin{abstract}
We explore the distribution in space and time of a single-photon excitation shared by a network of dipole-dipole interacting atoms that are also coupled to a common photonic field mode. Time-averaged distributions reveal partial trapping of the excitation near the initially excited atom. This trapping is associated with resonances of the excitation at crossing points of the photon-dressed energy eigenvalues of the network. 
The predicted \textit{photon-induced many-atom trapped excitation} (PIMATE) is sensitive to atomic position disorder which broadens the excitation resonances and transforms them to avoided crossings.  PIMATE is shown to allow highly effective and accurate sensing of \textit{multi-atom} networks and their disorder.
\end{abstract}

\maketitle
\section*{Introduction}

\bblk{Many-body quantum physics has made tremendous progress in recent years concerning manifestations of multipartite correlations/entanglement~\cite{toth_2005, calabrese_2009, bianchi_2022,li2018perfect,manmana,sodano,chiara} and their propagation~\cite{bose,  subra1,subrahmanyam_2006,zwick_2014}, the associated quantum information transfer and processing~\cite{kurizki_2015_pnas,georgescu_2014,bernien_2017}, as well as quantum phase transitions~\cite{coleman2005quantum,luca_2017_prl} and many-body localization via disorder~\cite{abanin2019colloquium,vosk2015theory,gogolin1982electron,shepelyansky_1994_prl}, to mention but a few directions. The highly complex character of quantum many-body systems has prompted the effort to simulate or probe their dynamics, experimentally and computationally~\cite{zhang2022quantum,asadi2020protocols,saffman2016quantum,tiarks2019photon,su2023rabi}. Yet realistic, analytically solvable, many-body quantum models are still scarce.} 

\bblk{Here we identify such a dynamical model for which analytical solutions are obtainable via generalization of the Wigner-Weisskopf approach~\cite{arnab_2017} to many-body systems, revealing the sensitivity of these solutions to spatial disorder in the system, i.e. to random deviations of atomic positions from regular spacing. The model concerns many atoms that are coupled to each other via the resonant dipole-dipole interaction (RDDI)~\cite{thiru,lehmberg_pra_1970,ritter2012elementary,kofman_kurizki} along with their individual strong coupling to a field mode in a cavity or a waveguide~\cite{paternostro2009solitonic,corzo2019waveguide,bardroff1995dynamical,PhysRevA.88.033830}. This single-mode coupling to the field mode is particularly strong when the atoms are excited to a high-lying electronic Rydberg state which can have a large dipole moment~\cite{cohen1998atom,lambropoulos2007fundamentals}. The outcome is consecutive absorption and re-emission of the microwave photon by different atoms that act as two-level atoms (TLAs) which are coupled to (are ``dressed" by) the field mode~\cite{cohen1998atom}.}

\bblk{The precision of such many-atom disorder sensing is statistically quantified by either the classical or the quantum Fisher information (FI)~\cite{helstrom1969quantum,paris2009quantum,victor} of diagnosing the sensitivity of a distribution to a parameter value. These crossings give rise to \textit{scattering resonance} that partially trap the excitation over long times. } 

\bblk{Although RDDI effects are prominent in many-body physics~\cite{paternostro2009solitonic,corzo2019waveguide,Gershon,defenu2023long,sheremet2023waveguide,drori2023quantum,opatrny2023nonlinear}, most treatments do not allow for the full dependence of the RDDI on the interatom distance $r$ which varies between $\sim 1/r^3$ for near-zone distances $r<< \lambda$ to $\sim 1/r$ for far-zone distances $r\gtrsim \lambda$, $\lambda$ being the resonant wavelength of the atomic transition~\cite{david}. Likewise, there has been little concern with the corresponding $r-$dependent radiative dissipation that inevitably accompanies RDDI in free space~\cite{shahmoon2016highly,lehmberg_pra_1970}. Finally, in photonic waveguides composed of Bragg-scattering elements, the $r-$dependence of RDDI coupling is \textit{profoundly modified near and inside photonic bandgaps}~\cite{shahmoon2011strongly,PhysRevA.72.043803,kofman1994spontaneous}, thus providing a powerful handle on many-atom dynamics, yet to be explored.}
 
Here we predict a hitherto unexplored effect which we dub \textit{photon-induced many-atom trapped excitation} (PIMATE). This effect consists of \textit{partial spatial trapping} (i.e, trapping accompanied by leakage) of a \textit{single excitation} shared by a network of RDDI coupled atoms that are simultaneously 
strongly interact with a photonic field mode (see Model). This partial trapping is associated with the resonances of the excitation.  \bblk{In a dissipationless (closed-system) scenario, these resonances arise at crossing points of eigenvalues~\cite{anderson_1958_pr,cohen1998atom}. These crossing points are an unfamiliar feature of the spectra of many-body but finite-size Hamiltonians. They are reminiscent of analogous features of infinite systems with long-range interactions that undergo phase transitions~\cite{castanos2012quantum,solinas2008dynamical}, notwithstanding profound differences between the systems (see Discussion). }

\bblk{In the present system, the crossing points occur at specific values of RDDI and, correspondingly, specific interatom distances for a given coupling strength of the atoms to the photonic mode (see Results).}
As a result, PIMATE is shown to be highly sensitive to the \textit{long-range (retardation)} characteristics of RDDI~\cite{shahmoon2016highly,shahmoon2011strongly,PhysRevA.72.043803,Gershon} and its \textit{disorder} in the many-atom network.  Unlike single-body Anderson localization (AL)~\cite{anderson1958absence,segev2013anderson,brandes2003anderson,abrahams201050,nagaoka2012anderson} or many-body localization (MBL)~\cite{gogolin1982electron,abanin2019colloquium,vosk2015theory}, disorder in the many-atom system \textit{weakens} PIMATE by broadening the excitation resonances and transforming the crossings into pseudo-crossings. 
\bblk{This property allows PIMATE to serve as a sensor of the many-atom disorder. The disorder can be sensed by non-demolition monitoring of the time-dependent and time-averaged excitation of the initially-excited atom, owing to the remarkable properties of PIMATE (see Discussion). }

The precision of such many-atom disorder sensing is statistically quantified by either the classical or the quantum Fisher information (FI)~\cite{helstrom1969quantum,paris2009quantum,victor} of diagnosing the sensitivity of a distribution to a parameter value. We show that the \textit{FI of estimating the disorder of the many-atom distribution diverges near the crossing-points}. This divergence corresponds to highly enhanced precision of mapping out the many-atom distribution and its disorder by merely a \textit{single measurement} of the excitation of a \textit{single (initially-excited) atom}. In contrast, without the photon-induced trapping that underlies PIMATE, as in open space, we would need to probe many atoms in RDDI-coupled disordered atom chains in open space and perform many more measurements of each atom in order to obtain similar information. Thus, PIMATE constitutes a novel, highly effective \textit{single-atom sensor} of \textit{multi-atom} disorder~\cite{david}. \bblk{This feature is the central result of our work (see Results, Discussion). }



\section*{Theory}
\subsection*{Model}

\label{sec1}
\begin{figure*}
\centering
\subfigure{%
  \includegraphics[width=0.4\textwidth]{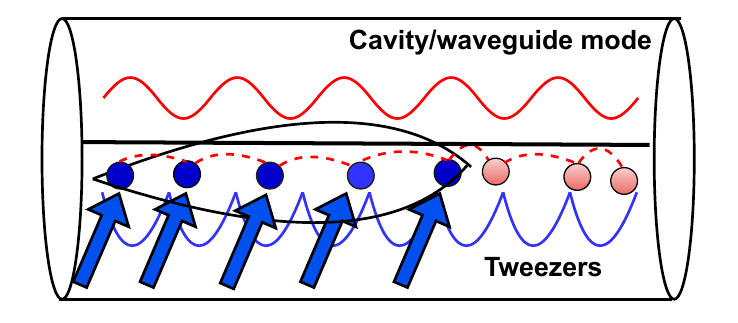}
  }
    \subfigure{%
  \includegraphics[width=0.48\textwidth, height=0.14\textheight]{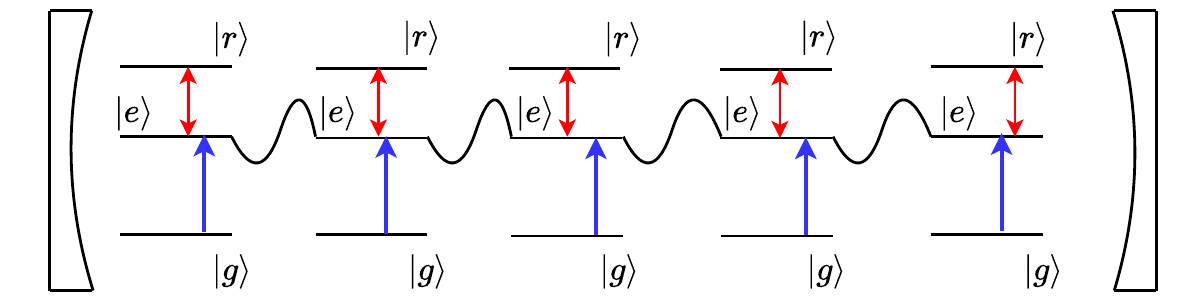}
  }
\caption{ \bblk{ (a) Schematic of ordered or disordered dipole-dipole interacting N-atom chain that is strongly coupled to a single-photon cavity/waveguide mode. The blue arrow denotes a laser beam focused on the chosen $N$ atoms that excites them to a state where they can interact with the field mode. (b) Schematic of three-level N Rydberg atoms in a microwave cavity, where the atoms are excited from $|g\rangle$ to $|e\rangle$ by a focused laser beam, from which they can be further excited to the Rydberg state $|r\rangle$ by a microwave photon in the cavity mode.} }
\label{fig:label1}
\end{figure*}

\begin{figure*}
\centering
   \subfigure[]{%
  \includegraphics[width=0.47\textwidth]{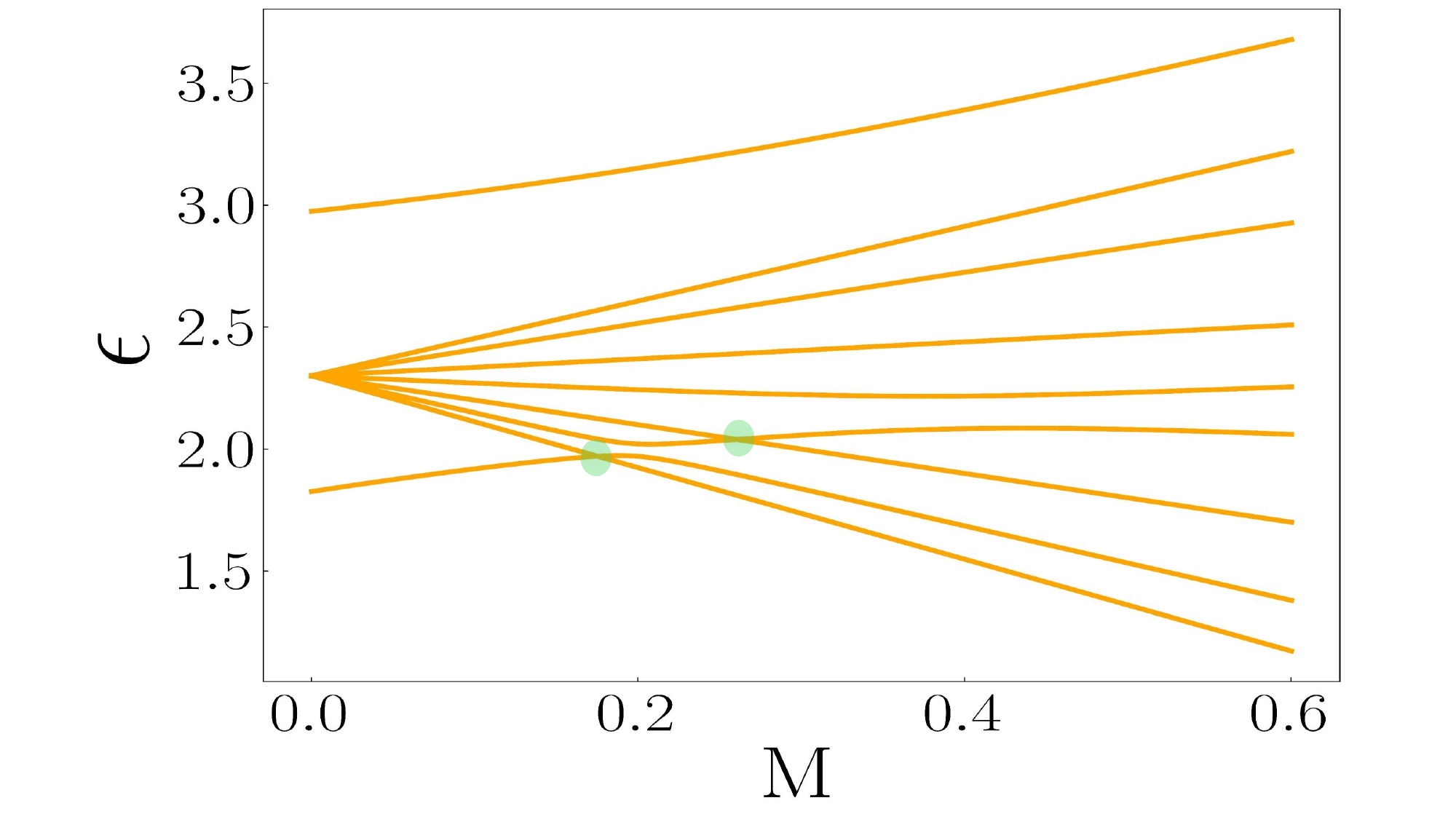}
  }
       \subfigure[]{%
  \includegraphics[width=0.48\textwidth]{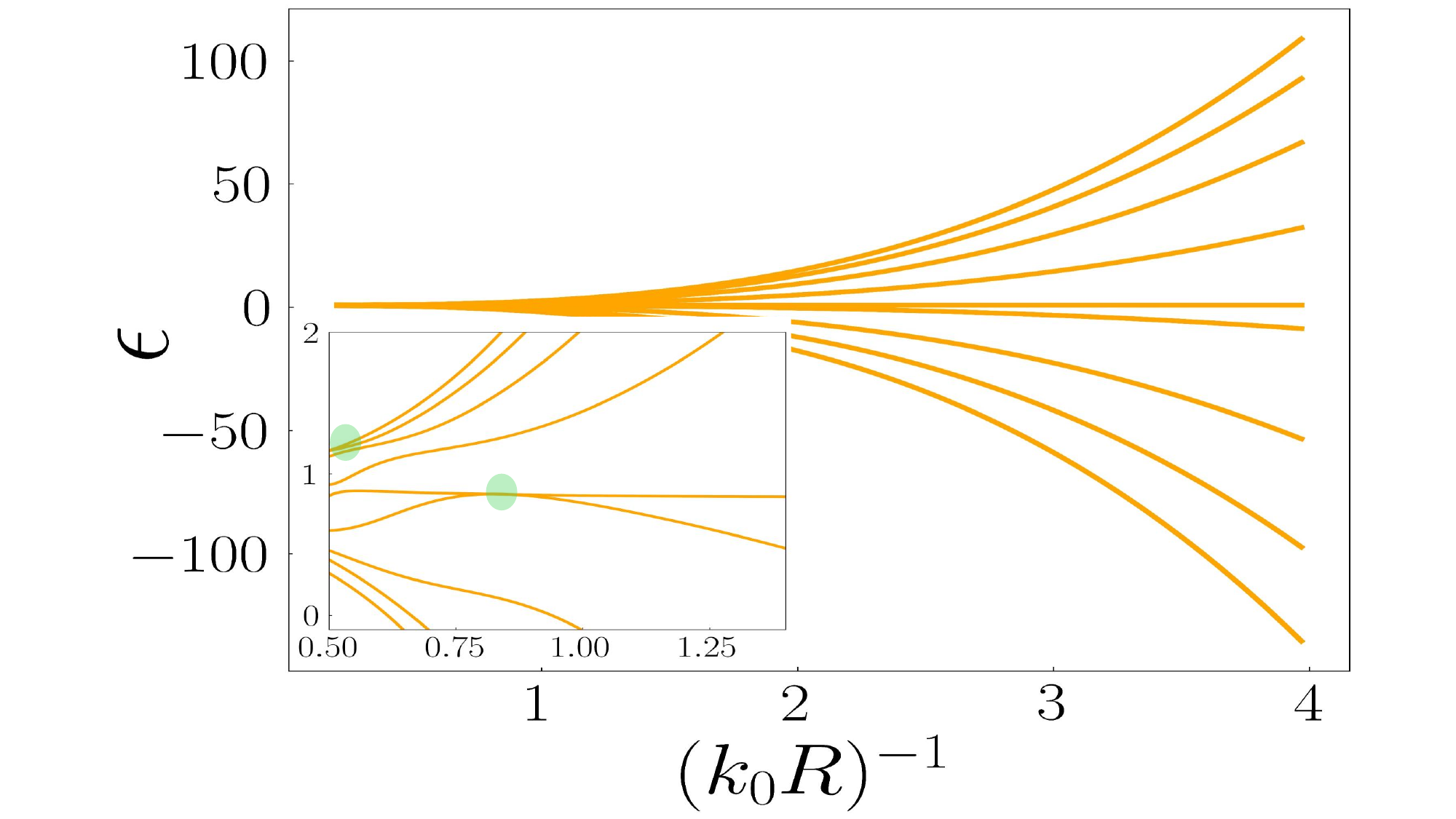}
  }
\caption{ \bblk{Single-excitation eigenvalues \bblk{($\epsilon$)} for an open chain of $N=8$ atoms as a function of the RDDI strength $M(R_{jj'})$ (\bblk{in units of the atomic radiative rate $\gamma$}) and fixed atom-cavity coupling $\kappa_j$ \bblk{(same units)}, assuming negligible dissipation (which is suppressed by a high-Q cavity or a photonic bandgap waveguide). The system is described by Hamiltonian parameters: $\omega - \omega_0 = 0.2 \gamma$, $\kappa_j = 0.2 \gamma$ where $\omega$ is the mode frequency and \bblk{$\omega_0$ is the $|e\rangle-|r\rangle$ transition frequency, and $M(R_{ij})$ where $ R_{ij}= |X_i-X_j|$ are the atomic positions, here regularly spaced without disorder.} (a) Nearest-neighbor (NN) RDDI \bblk{where $R\equiv R_{ij} = |X_i-X_{i\pm 1}|$} (b) Same, for long-range non-nearest-neighbor (NNN) RDDI as a function of the interatomic separation $k_0 R\equiv k_0 R_{ij}$ \bblk{the RDDI length scale being the $\lambda_0 =2\pi/k_o$, $k_0=\omega_0/c$}. The crossing points and the pseudo-crossing points of eigenvalues on all plots are marked by circles \bblk{(see explanations in text)}.}}
\label{fig:label1v1}
\end{figure*}
\begin{figure*}
  \begin{center}  
    \subfigure[]{%
  \includegraphics[width=0.48\textwidth]{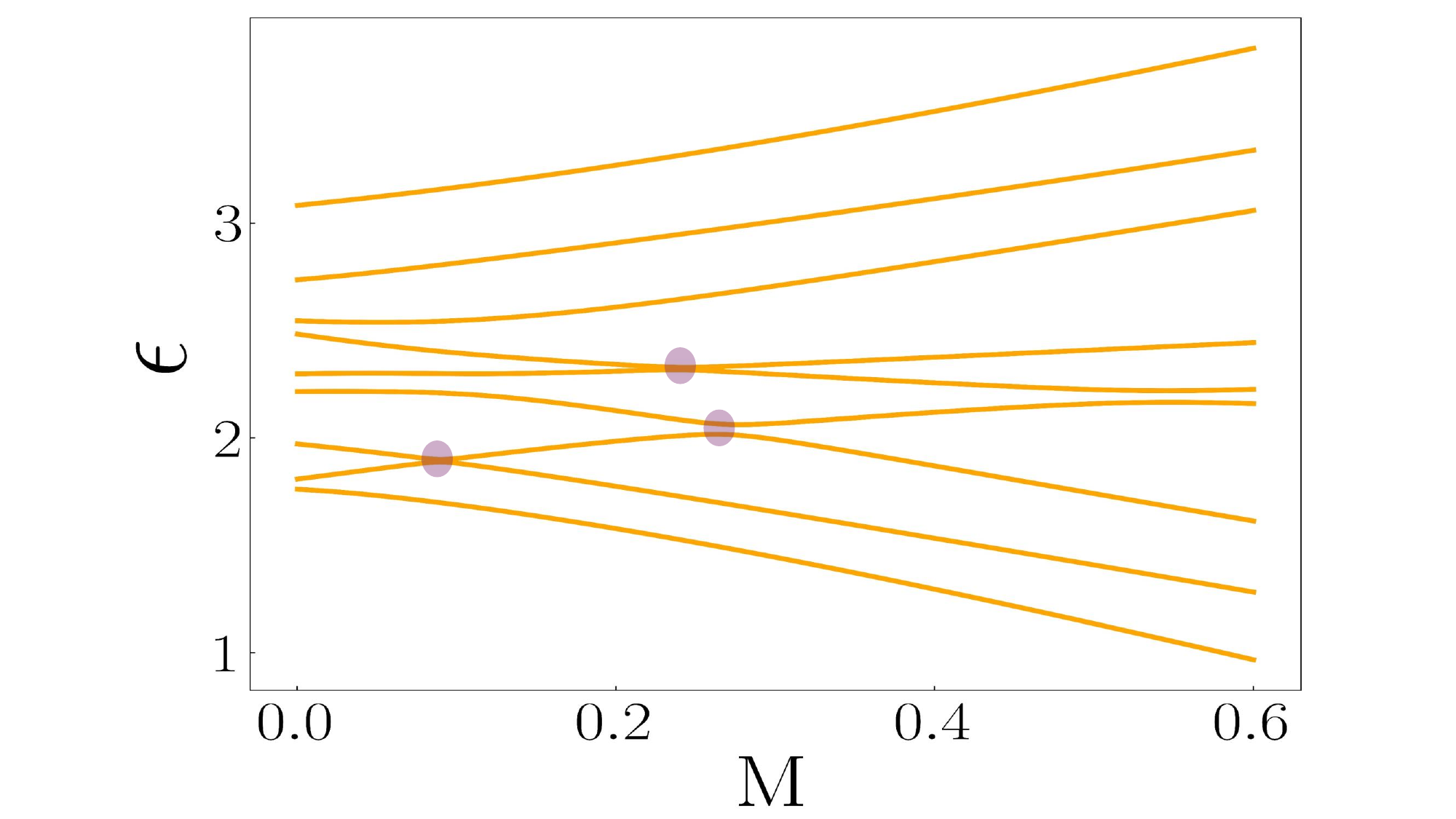}
  }
\subfigure[]{%
  \includegraphics[width=0.48\textwidth]{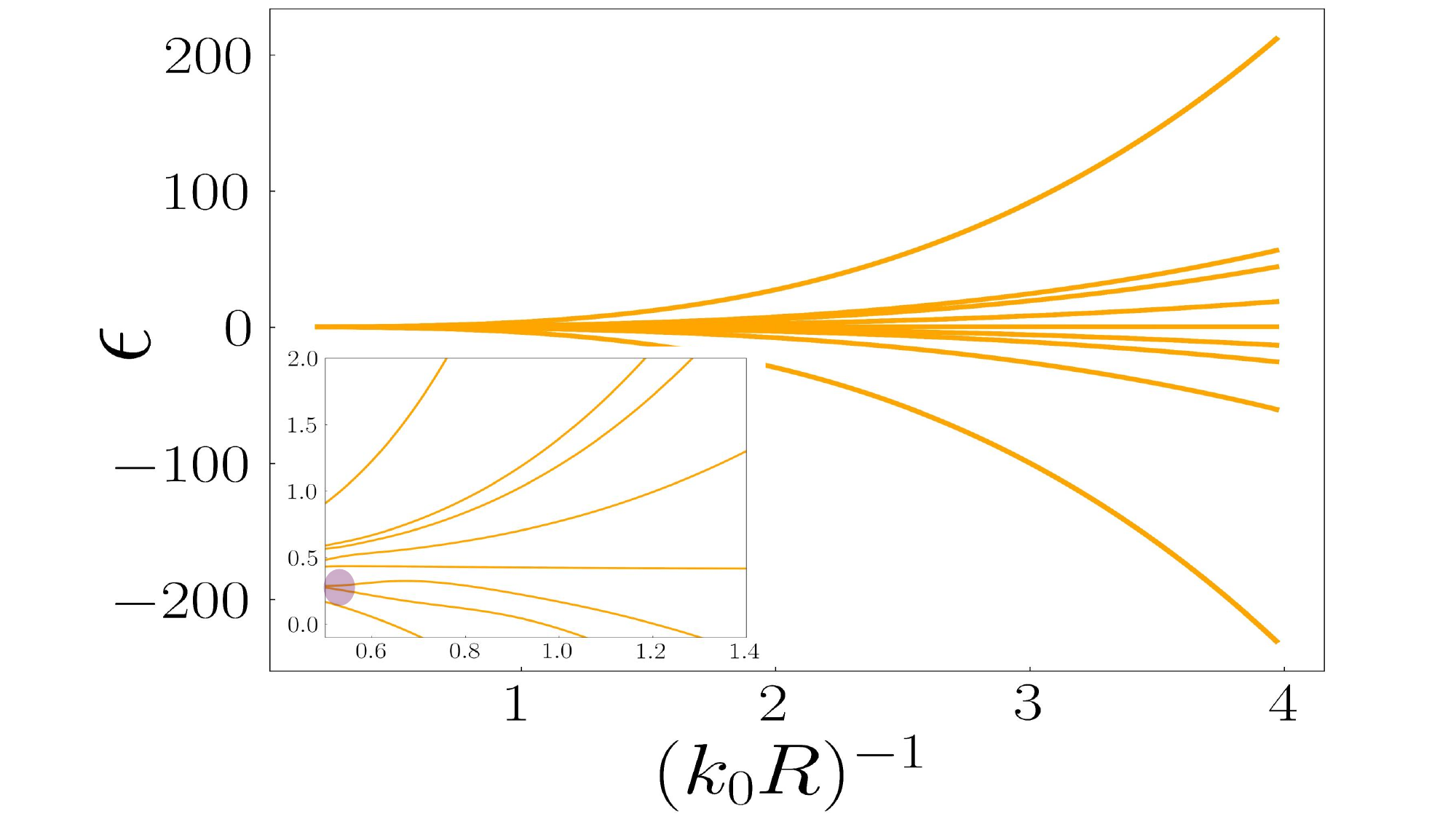}
  }\\
   \subfigure[]{%
  \includegraphics[width=0.42\textwidth]{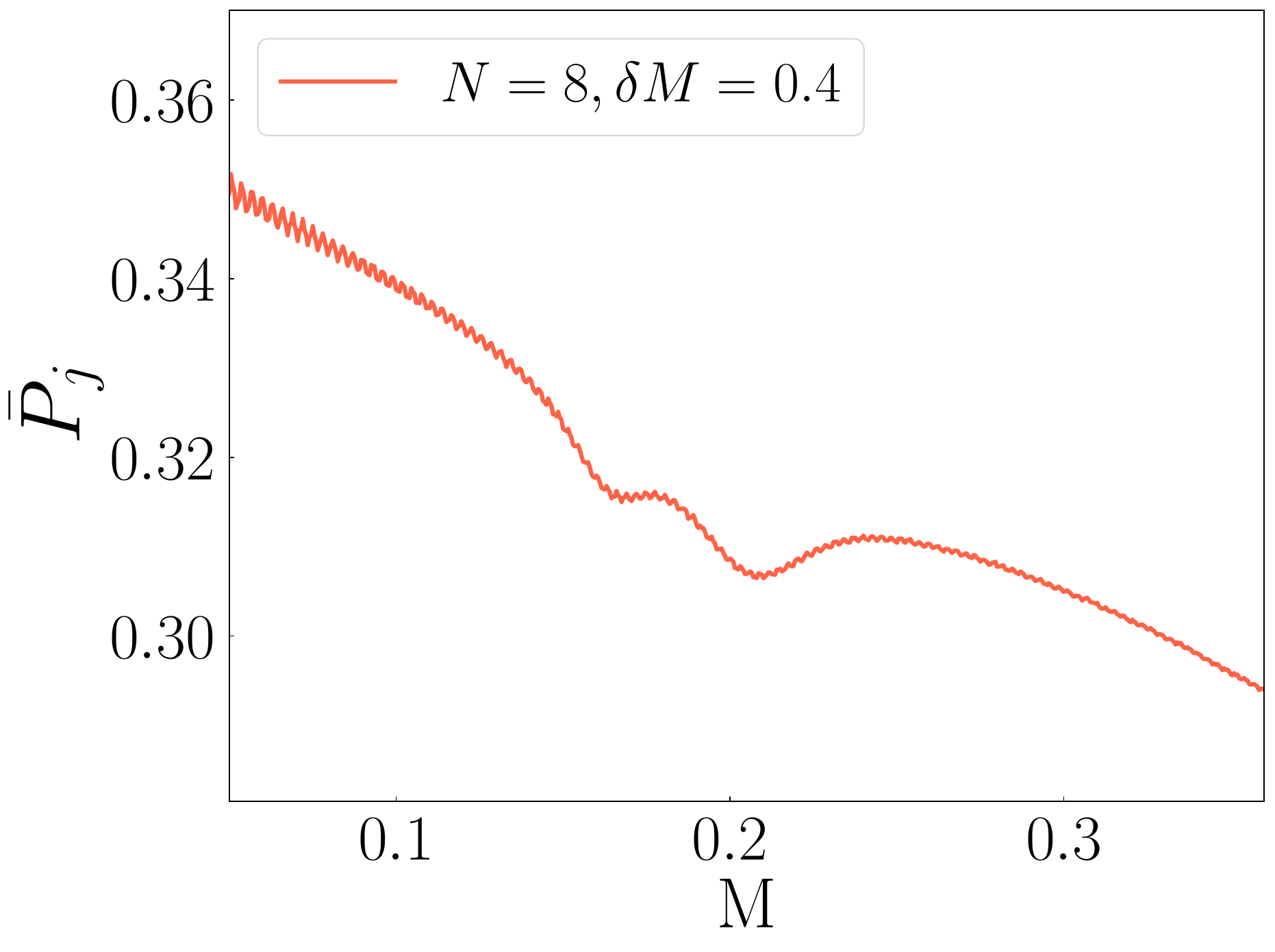}}
      \subfigure[]{%
  \includegraphics[width=0.4\textwidth]{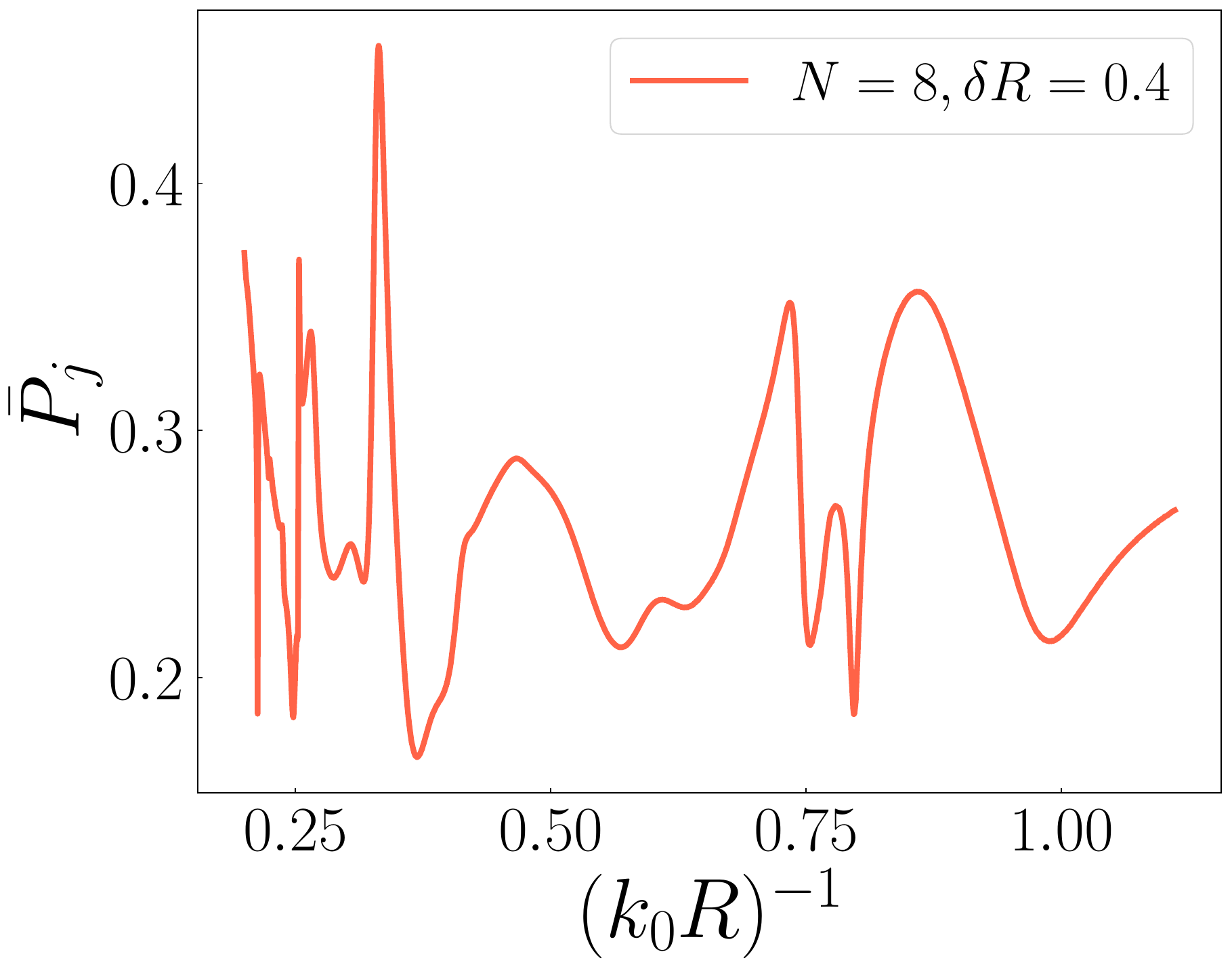}}
\end{center}
      \caption{{\bblk{(a), (b) Eigenvalue \bblk{($\epsilon$)} dependence on \bblk{RDDI, $M (R_{ij})$ (Eq. (\ref{eqnnnnnnnn11})) on its argument $k_0R$},  for the same parameters as in Fig.~\ref{fig:label1v1}, but in the presence of off-diagonal atom-position randomness with standard deviation \bblk{$\delta X_j/X_j$}= 0.4. The crossings and the pseudo-crossings are marked by circles. (c) Resonances of the time-averaged excitation probability \bblk{$\bar{P}_j$ (to state $|r_j\rangle$) of the initially-excited jth atom} (normalized to 1) as a function of $M$ \bblk{(in units of $\gamma$)} with negligible dissipation for the NN interaction. The plot indicates that trapping vanishes with high randomness \bblk{$\delta X_j$ as above}, (d)  Same, for the long-range NNN interaction with \bblk{the same randomness} as a function of $k_0 R$, yielding partial trapping (high peaks). The  Hamiltonian parameters are \bblk{as in Fig.~\ref{fig:label1v1}, and the averaging time is $T =10000/\gamma$}.}}}
      \label{fig:pole_N=4_RR}
  \end{figure*}

We consider spatially distributed atoms that form a chain and are coupled to a field mode of a waveguide or a cavity (Fig.~\ref{fig:label1}(i),(ii)). As a possible experimental scenario, we may envisage atoms with levels $|g\rangle$, $|e\rangle$, $|r\rangle$ placed in a microwave cavity (or waveguide) that is resonant with the $|e\rangle \rightarrow |r\rangle$ transition between the high-lying electronic (Rydberg)~\cite{cohen1998atom,lambropoulos2007fundamentals} states $|e\rangle$ and $|r\rangle$ (Fig.~\ref{fig:label1}b). Rydberg-state transitions are chosen here because they have large transition dipole moments, allowing for strong coupling to the field and large RDDI~\cite{zhang2022quantum,asadi2020protocols,saffman2016quantum,tiarks2019photon,su2023rabi,drori2023quantum}. 

\bblk{Let us assume that all the atoms in the system are initially in the ground state $|g\rangle$ that is resonant with the $|g\rangle \rightarrow |e\rangle$ transition. A laser is focused onto $N$ atoms and transfers all of them to the $|e\rangle$ state via a short $\pi$-pulse, which ensures full population transfer~\cite{lambropoulos2007fundamentals}. This \textit{abrupt laser-induced quench} exposes the $N$ atoms in state $|e\rangle$ to resonant excitation by the microwave cavity/waveguide field mode that has initially one photon.  After the microwave photon is absorbed by one of these atoms, the excitation can further propagate along the chain via the combination of RDDI and strong single-atom coupling to the field mode, that is resonant with the $|e\rangle \rightarrow |r\rangle$ transition. The outcome is consecutive absorption and re-emission of the microwave photon by different atoms that act as two-level atoms (TLAs) which couple to (are ``dressed" by) the field mode~\cite{cohen1998atom}. The combined $N-$atom chain plus the field mode are thus confined to the sector of \textit{single excitation}.}

\bblk{As shown below, the excitation spectrum of the atoms exhibits crossings when different energy eigenvalues are plotted versus the interatom separations. These crossings give rise to \textit{scattering resonances} that trap the excitation over long times. We find these crossings to be highly sensitive to disorder in the chain, i.e. to random deviations from the prescribed atomic positions.}

The \textit{effective Hamiltonian}\bb{~\cite{Gershon}} of this combined multi-atom-field system \textit{following the quench} is 
 \begin{eqnarray} \label{effect12}
    &&H =\hbar \omega a^\dagger a + \hbar \sum^N_{j=1} \omega_0 \vert r_j \rangle \langle r_j\vert  \nonumber\\
    && + \hbar \sum^{N}_{i,j=1; i<j} M_{ij} \left(\vert e_i r_{j} \rangle \langle  r_{i} e_{j} \vert + H.c.\right) 
  +\hbar \sum^N_{j=1} \left(\kappa_j a \vert r_j\rangle \langle e_j\vert+ H.c.\right).\nonumber\\
\end{eqnarray}
Here $\vert r_j \rangle$ and $\vert e_j \rangle$ constitute the upper and lower states, respectively, separated by the energy $\hbar \omega_0$, of the effective $j-$th TLA; $M_{ij}$ is the self-energy matrix element that expresses the RDDI strength between the $i$ and $j$ TLAs, $\kappa_j$ is the coupling strength of the $j$-th TLA with the near-resonant cavity mode at the frequency $\omega \approx \omega_0$, being the field mode characterized by annihilation operator $a$.
\bblk{The resonant photon wavelength $\lambda_0$ is the characteristic length scale of the RDDI so that the separation $R_{ij}<< k^{-1}_0$, where $k_0= 2\pi/\lambda$ corresponds to near-zone interaction of the form }
\begin{subequations}
    \begin{equation}
        M_{ij} \sim (k_0 R_{ij})^{-3}.
    \end{equation}
\bblk{By contrast, for $R_{ij}$ beyond the near-field zone, RDDI must be described by its full separation-dependent real self-energy matrix~\cite{thiru,defenu2023long,sheremet2023waveguide,ficek2005quantum,lehmberg_pra_1970}. }
\bblk{The atomic dipole can be oriented perpendicular (denoted by $(\Pi)$) or parallel (denoted by $\Sigma$) to the multi-atom chain. Here we consider the $\Pi$ orientation of the atomic dipole to the multi-atom chain axis for which the real part of the self-energy matrix $\Re(M_{ij})$ has the form~\cite{lehmberg_pra_1970} (see SI)}  
\begin{eqnarray}\nonumber
 \text{Re}(M_{ij}) \equiv M(R_{ij}) = \gamma J_{\Pi} (k_0R_{ij}),
\end{eqnarray}
 
\begin{eqnarray} \nonumber \label{eqnnnnnnnn11}
&& J_{\Pi} (k_0 R_{ij}) = \frac{3}{4}\left[\frac{\cos (k_0 R_{ij})}{(k_0 R_{ij})^3} + \frac{\sin(k_0 R_{ij})}{(k_0 R_{ij})^2} - \frac{\cos (k_0 R_{ij})}{(k_0 R_{ij} )} \right].\\
\end{eqnarray}
\end{subequations}
\bblk{which varies as $\cos (k_0R_{ij})/k_0 R_{ij}$ at $R_{ij}\gtrsim \lambda_0$ (the far zone). The magnitude of RDDI proportional to $\gamma$ is the single-atom radiative rate at the $|r_j\rangle \rightarrow |e_j\rangle$ transition in free space~\cite{ficek2005quantum,sheremet2023waveguide,defenu2023long,thiru,lehmberg_pra_1970}.}


In general, one should also account for the \textit{dissipative}, radiative effect associated with the \textit{imaginary} self-energy Im$M_{ij}$ that complements the real self-energy matrix Re$M_{ij}$~\cite{ficek2005quantum,sheremet2023waveguide,defenu2023long,thiru,lehmberg_pra_1970}. The dissipation analysis is relegated to the SI. In the main text, we assume \textit{negligible radiative dissipation} as in a high-Q cavity or in a photonic bandgap structure (Bragg grating embedded in a waveguide)~\cite{gogolin1982electron,abanin2019colloquium,vosk2015theory,kurizki_1990_pra}. 

\begin{figure*}
  \begin{center}
    \subfigure[]{%
  \includegraphics[width=0.45\textwidth]{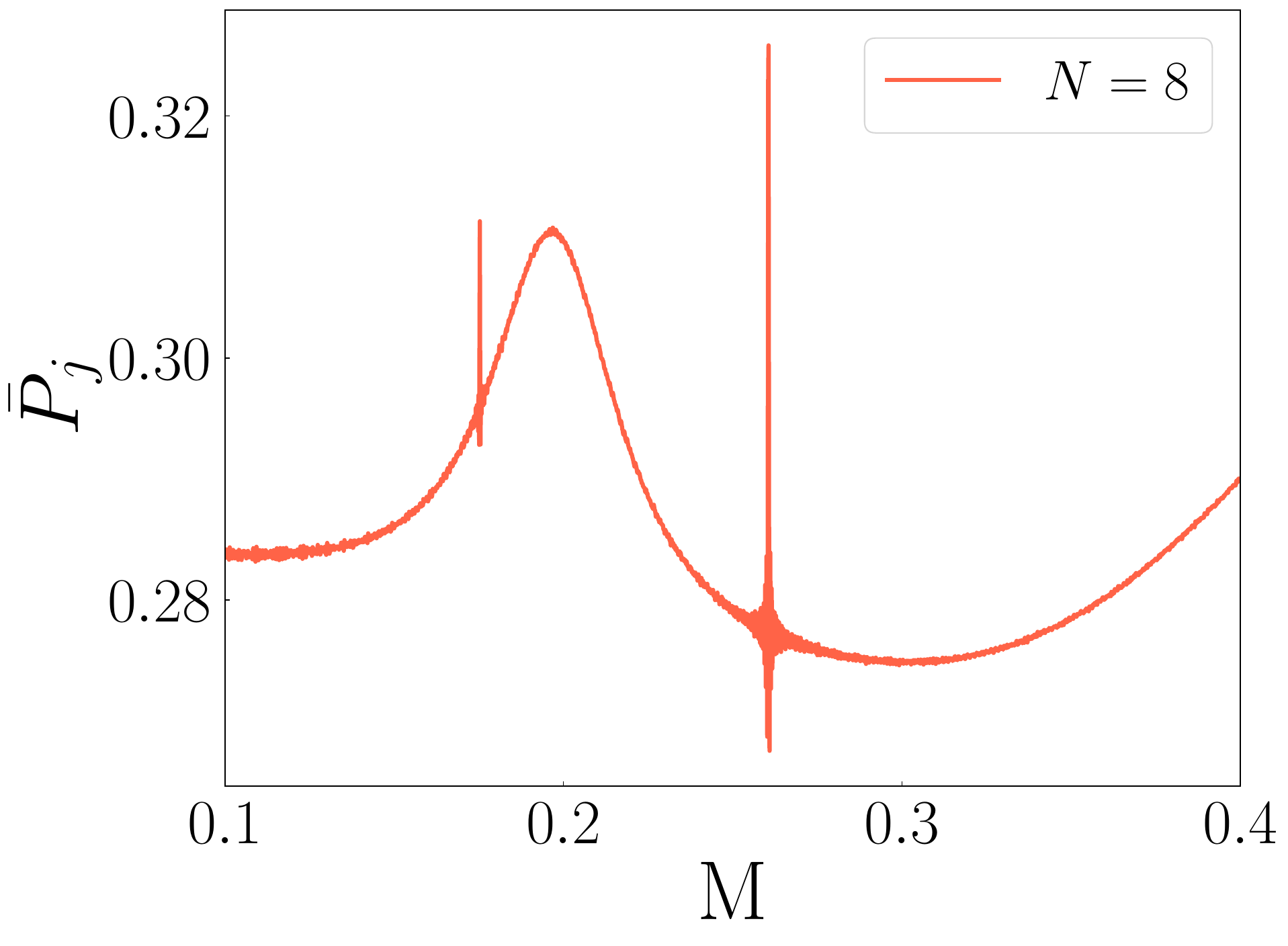}}
      \subfigure[]{%
  \includegraphics[width=0.45\textwidth]{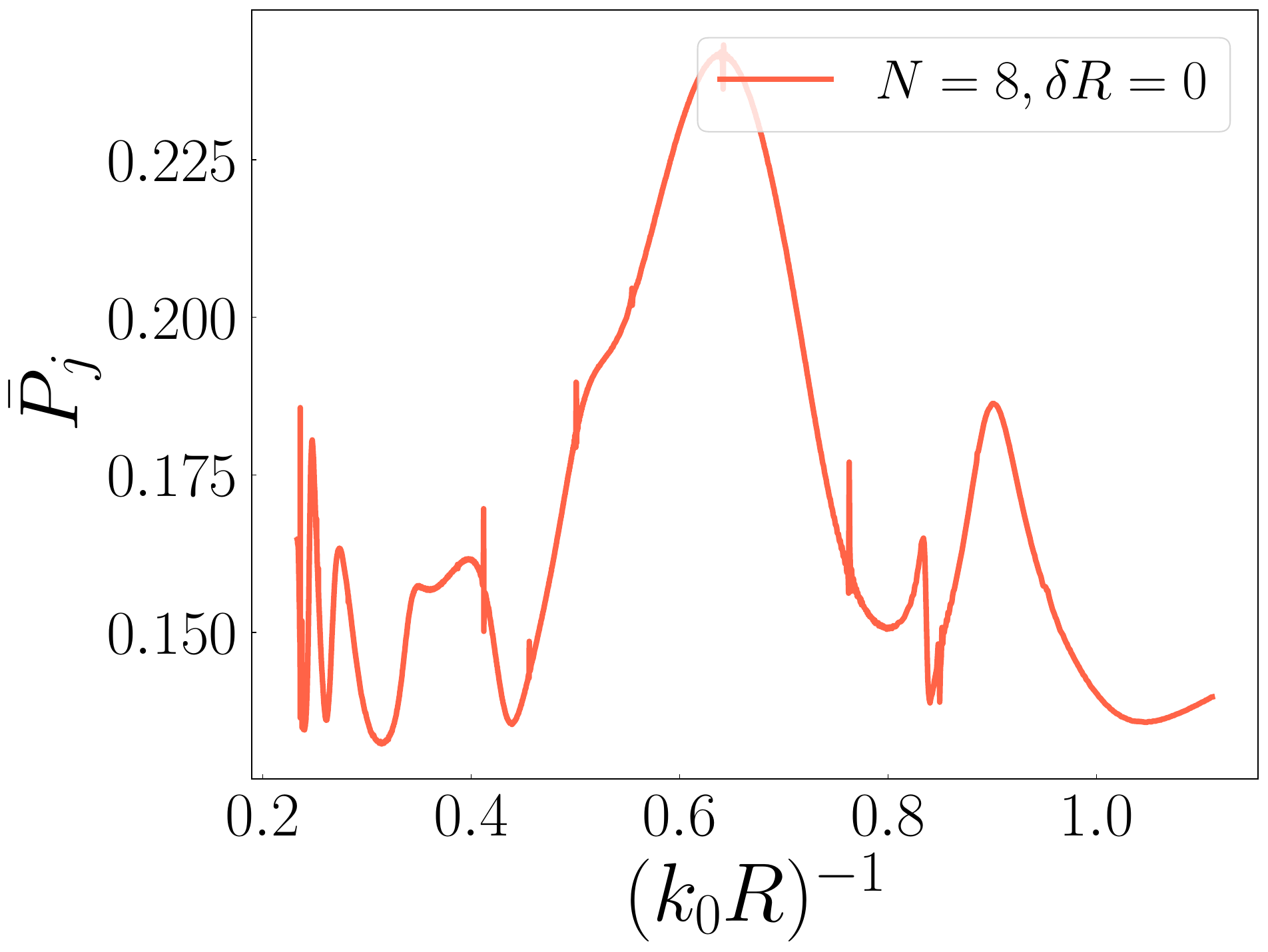}}
\end{center}
      \caption{(a)\bblk{ Same as in Fig.~\ref{fig:pole_N=4_RR}(c), \ref{fig:pole_N=4_RR}(d) but without position randomness}
      for the nearest-neighbor (NN) interaction.  Partial trapping \bblk{of the excitation} occurs at the crossing point of the eigenvalue spectrum in perfectly periodic chains, (b)  Same, as a function of $k_0 R$ for the \textit{long-range} (NNN) interaction.  The  Hamiltonian parameters \bblk{and averaging time are as in Fig.~\ref{fig:label1v1}, \ref{fig:pole_N=4_RR}}. }
      \label{fig:prob_WOR}
  \end{figure*}

\begin{figure*}
  \begin{center}
  \subfigure{%
  \includegraphics[width=0.45\textwidth]{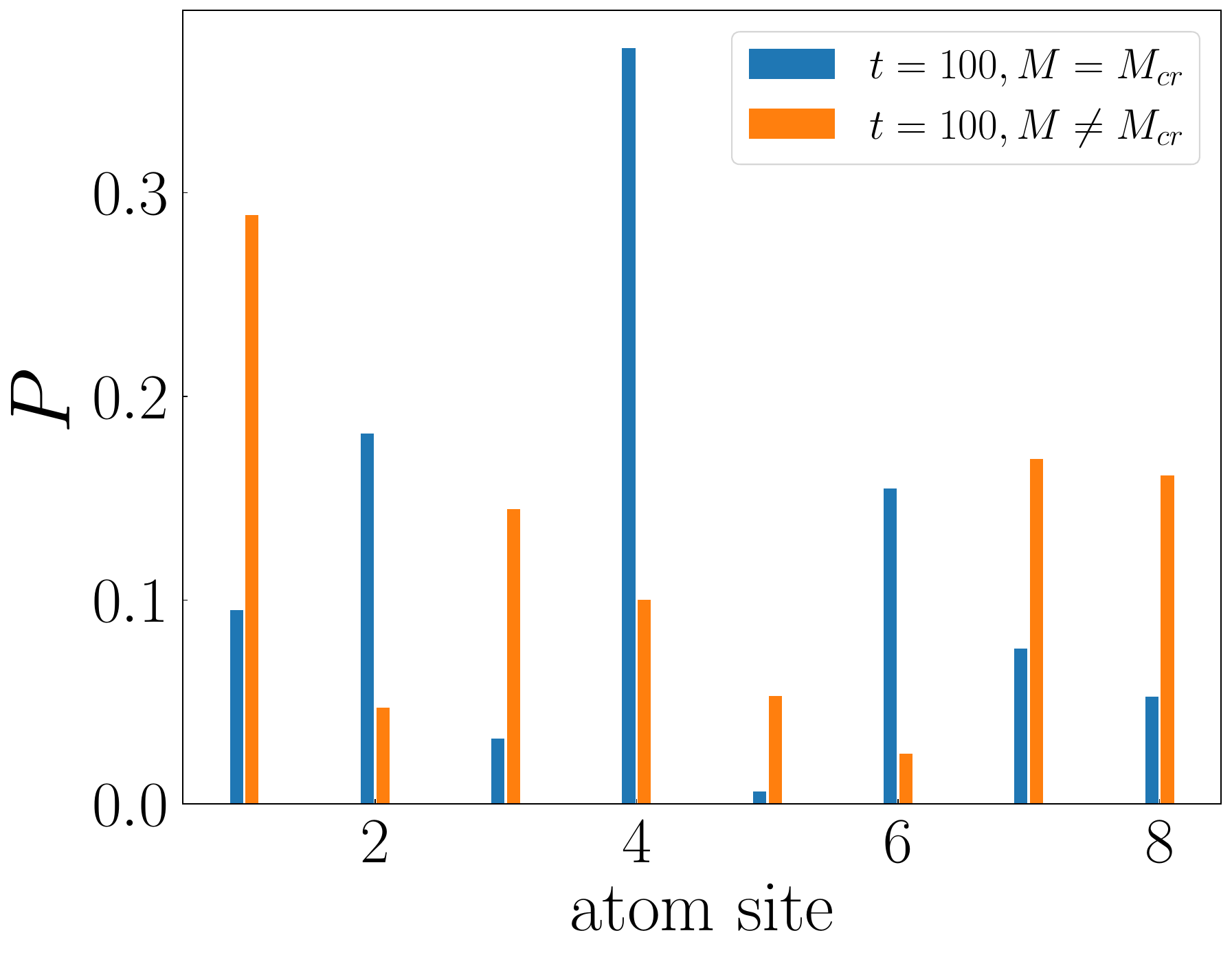}}
 \subfigure{%
  \includegraphics[width=0.45\textwidth]{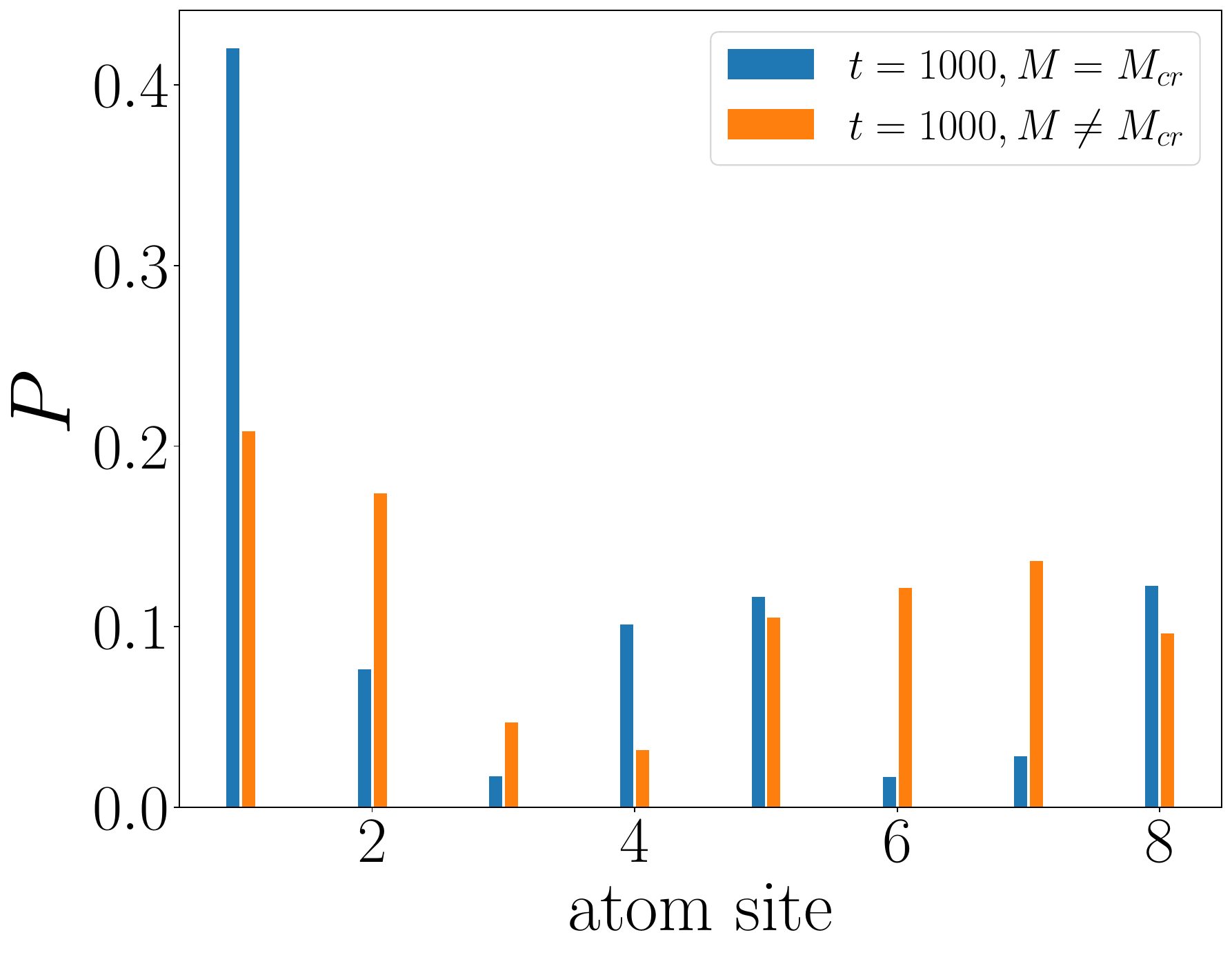}}
  \end{center}
      \caption{\bblk{Time-resolved excitation probability of individual atoms to state $|r_j\rangle$} at different times in the NN case of Fig.~\ref{fig:prob_WOR}a at the crossing/resonance point $M (k_0 R)=M_{cr}$ (in blue) and at a non-resonance point $M\neq M_{cr}$ (in orange). The  Hamiltonian parameters are as in Fig. \ref{fig:label1v1}. Partial (incomplete) trapping of the excitation at the site of the initially excited atom (here atom 1) is seen to occur at long times (right panel) \textit{only at the crossing point} in the NN case.} 
      \label{Time_dynamics_M}
  \end{figure*}

\begin{figure*}
  \begin{center}
  \subfigure[]{%
  \includegraphics[width=0.45\textwidth]{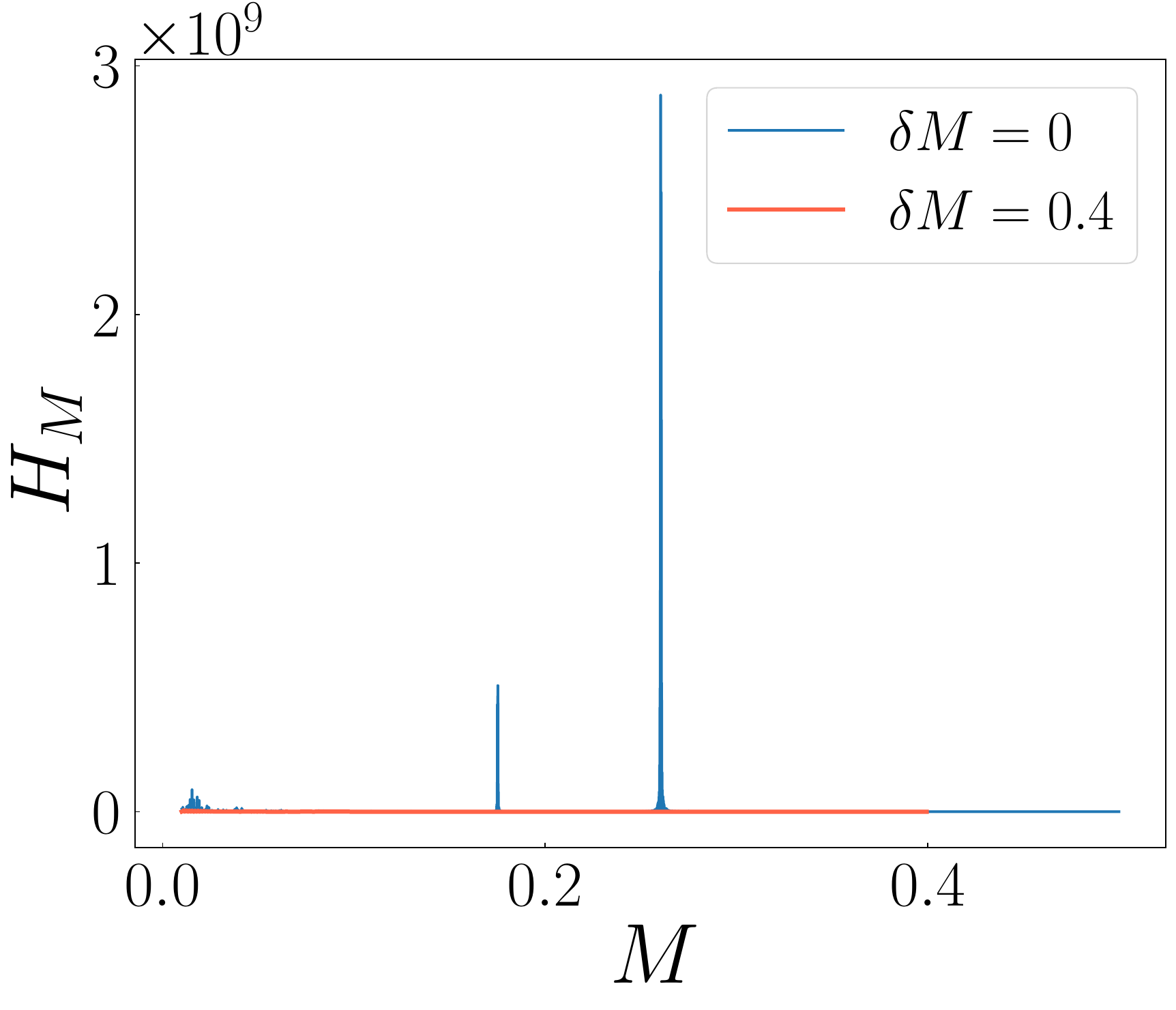}}
 \subfigure[]{%
  \includegraphics[width=0.45\textwidth]{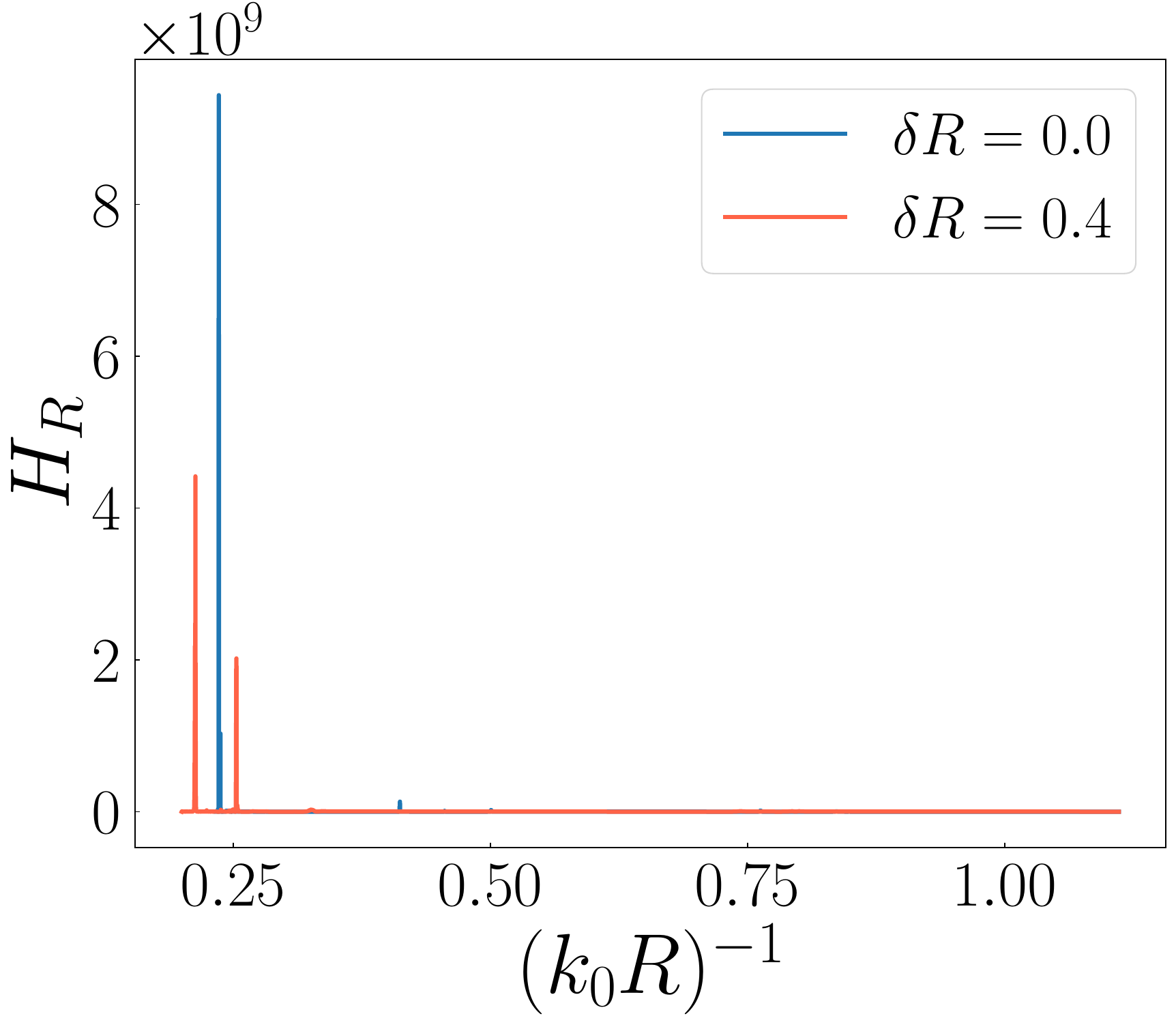}}
  \end{center}
      \caption{(a) Variation of the Fisher information without dissipation for \textit{nearest-neighbor} (NN) interacting $N=8$ atoms as a function of $M$ (\bblk{as in Fig.~\ref{fig:label1v1}-\ref{fig:prob_WOR}}) in the presence of off-diagonal randomness (red solid line) as in Fig.~\ref{fig:prob_WOR} and without randomness (blue solid line). (b) Same, \textit{for long-range NNN interaction}  for negligible dissipation as a function of $1/(k_0 R)$ 
      in the presence of off-diagonal randomness as in Fig.~\ref{fig:prob_WOR} (red solid line) and without randomness (blue solid line). The  Hamiltonian parameters  are as in Fig.~\ref{fig:label1v1}-\ref{fig:prob_WOR}.}
      \label{fig:fisher_N=4,8_WOR}
  \end{figure*}

\begin{figure*}
  \begin{center}
  \subfigure[]{%
  \includegraphics[width=0.46\textwidth]{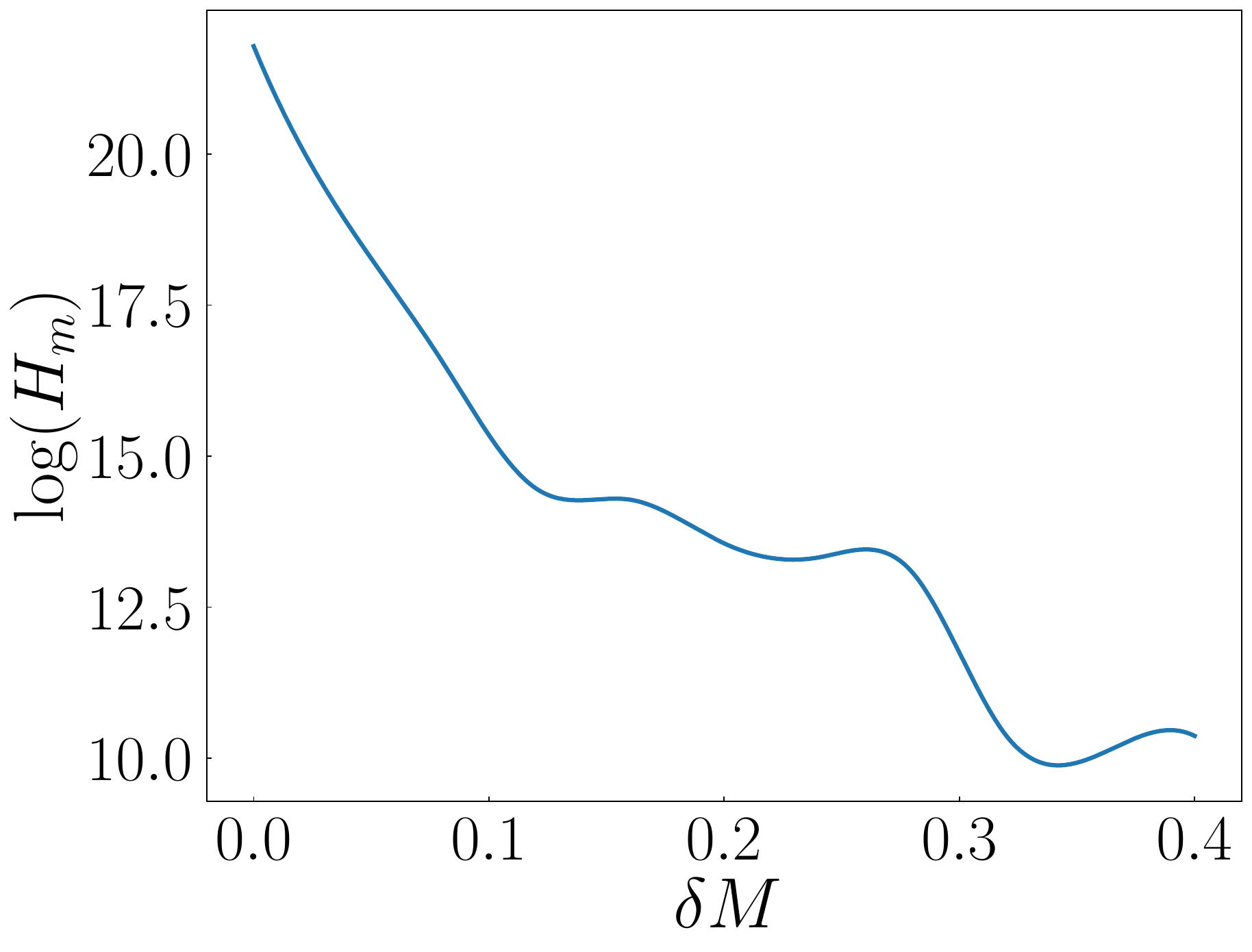}}
 \subfigure[]{%
  \includegraphics[width=0.45\textwidth]{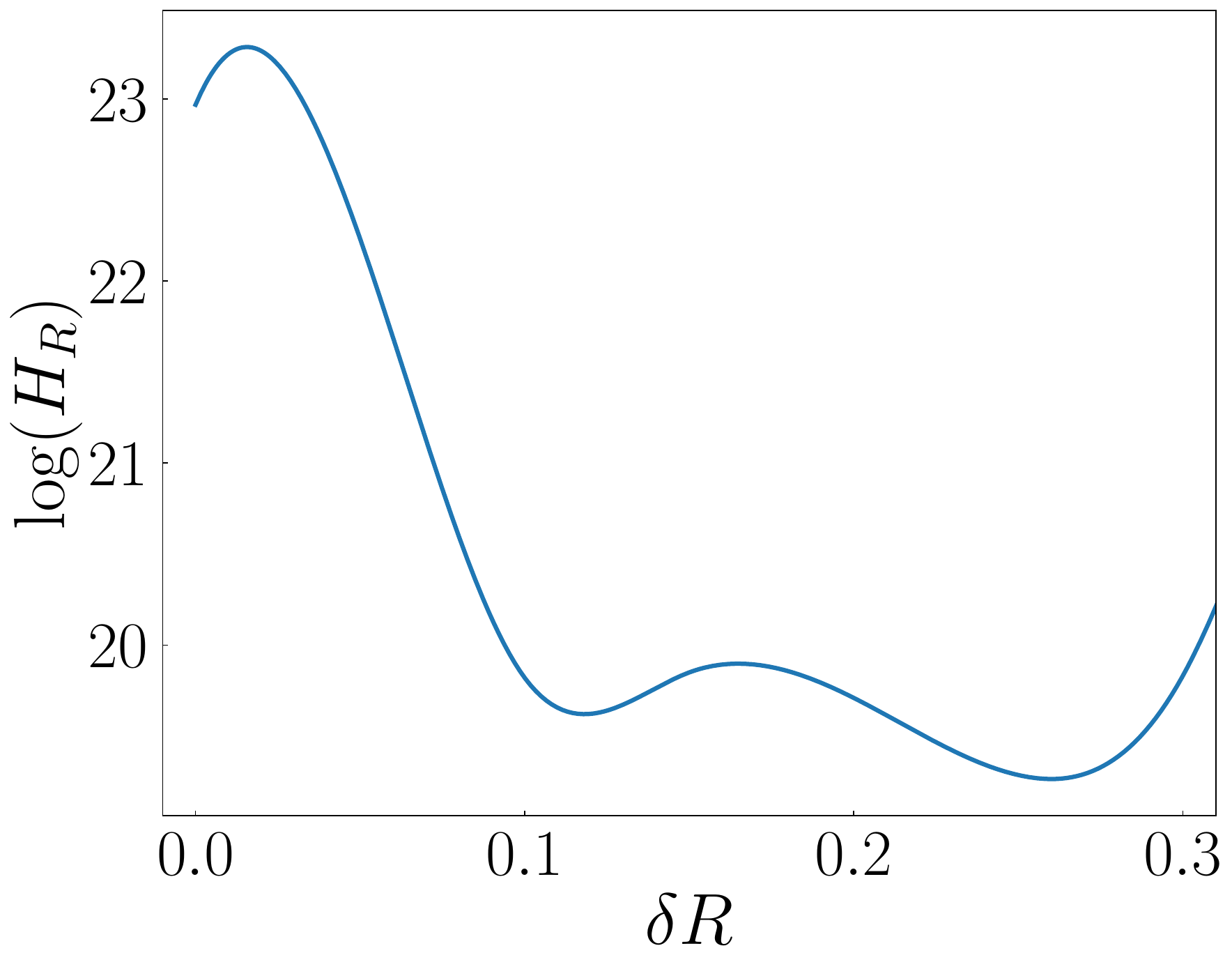}}
  \end{center}
      \caption{(a) The variation of the logarithmic Fisher information for $N=8$ as a function of randomness in $M$ for the NN interaction. (b) Same for the long-range NNN interaction.}
      \label{fig:fisher_N=8_WVR}
  \end{figure*}

\vspace{2ex}
\bblk{\subsection*{Analysis}}
 The most general time-dependent state\bb{~\cite{lehmberg_pra_1970,kofman_kurizki}} of the combined field-multi-atom system \eqref{effect12} in the \textit{single-excitation sector} is,
\begin{eqnarray}\label{eqn22}
\vert \psi(t) \rangle = \sum^N_{j=1} \alpha_j (t) \vert e,\dots,r_j, \dots e, \, 0 \rangle + \beta (t) \vert e...e, 1\rangle 
\end{eqnarray}
Here $0$ and $1$ denote the two possible photon numbers in the field mode. The  time-dependent amplitudes $\alpha_j (t)$ and $\beta (t)$ in Eq.~\eqref{eqn22} can be obtained from the Schr\"odinger equation 
 $\ket{\dot{\psi}(t)}=-iH \ket{\psi(t)}$ governed by Hamiltonian (1),
which yields a set of coupled differential equations for $\alpha_j (t)$ and $\beta (t)$.

Here we analyze the exact solution of the coupled equations by casting them into a set of coupled algebraic equations
in the Laplace domain\bb{~\cite{kofman1994spontaneous}}. The Laplace-transformed excitation amplitudes have the form, 
\begin{eqnarray}
 &&s\hat{\beta} (s) = -i \omega \hat{\beta}(s) - i \sum_j \kappa^*_j \hat{\alpha}_j(s) + \hat{\beta}(0), \nonumber\\
&&  s \hat{\alpha}_j(s) = -i {\omega_{0}} \hat{\alpha}_j(s) - i\kappa_j \hat{\beta}(s) - i \sum_{j'} M_{jj'} (R_{jj'}) \hat{\alpha}_{j'}(s) + \hat{\alpha}_j(0). \nonumber\\
\end{eqnarray}
Substituting the solution for $\hat{\beta} (s)$, under the assumption that $\hat{\beta} (0)=0$ (initially excited $j$th atom and empty cavity),  we can write the following solution for the \textit{vector} of the many-atom excitation amplitudes, $\hat{\boldsymbol{\alpha}}(s) \equiv (\hat{\alpha}_1(s),\hat{\alpha}_2(s),...,\hat{\alpha}_N(s))^T$:

\begin{subequations}
\begin{eqnarray}
    \hat{\boldsymbol{\alpha}}(s) =  \mathbf{A^{-1}}(s)\hat{\boldsymbol{\alpha}}(0) = \vert \mathbf{A}(s)\vert^{-1} \rm{adj}(\mathbf{A}(s))\hat{\boldsymbol{\alpha}}(0).
    \label{eq:soln}
\end{eqnarray}
Here, under the assumption of \textit{negligible dissipation} discussed above, the matrix elements have the form

 \begin{equation}
        \mathbf{A}_{jj'}(s)  =  (s + i{\omega_{0}}) \delta_{jj'} + \frac{\kappa_j \kappa^*_{j'}}{(s+i\omega)} + iM_{jj'} (R_{jj'}) (1-\delta_{jj'}).
\end{equation}
\end{subequations}


Imaginary roots of the polynomial obtained from $\vert \mathbf{A}(s)\vert$ by setting $s = \epsilon$ provide us with the actual spectrum of eigenvalues $\epsilon$. If $\{\epsilon\}$ are the solutions to $\vert \mathbf{A}(\epsilon)\vert = 0$, then $\prod_{m}(s-\epsilon_{m})^{m} = 0$; where the index $m$ denotes the order of the pole $\epsilon_{m}$. 
The solution to $\hat{\boldsymbol{\alpha}}(s)$ can be then expressed as
\begin{eqnarray}\label{eqnresi}
 \hat{\boldsymbol{\alpha}}(s) =   \frac{\rm{adj}(\mathbf{A}(s))\boldsymbol{\alpha}(0)}{\prod^l_{m=1}(s-\epsilon_{m})^{m}} =  \sum^l_{m=1} \sum^{m}_{j=1} \frac{\boldsymbol{r}_{m, j}}{(s-\epsilon_{m})^{j}},  
\end{eqnarray}
\bblk{where $r_{m,j}$ is the residue corresponding to the pole $\epsilon_{m}$ of order $j$.}

\bblk{The solution \eqref{eqnresi} has been obtained by the technique of partial fraction expansion of proper rational functions 
which can have different kinds of poles: Simple poles that are located far from critical points of the energy or higher-order poles located at critical points. The Laplace transform of the latter poles is divergent, \bblk{corresponding to oscillations (imaginary poles)} in the time domain. }

\bblk{Any crossing of energy eigenstates corresponds to higher-order poles. This crossing in turn gives rise to \textit{exponential divergence} in the residue, which determines the time-dependence of the associated population, alternating between growth and decay~\cite{mitrinovic1984cauchy} (see details in SI).} 

\bblk{Here, in the absence of dissipation the evolution described by solution~\eqref{eqnresi} is unitary, hence the inverse Laplace transform of $\hat{\boldsymbol{\alpha}}(s)$, i.e, the vector $\hat{\boldsymbol{\alpha}} (t)$ is oscillatory and undergoes recurrences. However, for $N \gg 1$ this evolution is analogous to that of a discrete level coupled to a dense energy band~\cite{akulin2012intense}: upon averaging the dynamics of $\alpha (t)$ over a long time interval, the initially populated level exhibits complex behavior}
of the time-averaged \textit{excitation probability $\Bar{P}_j$ of the initially excited atom} on the averaging time $T$. It is exponentially \bblk{decaying and partly oscillatory if there are only simple poles. Then} $\Bar{P}_j \propto e^{T(\lambda^*_{m'} + \lambda_{m})}$ where  the roots $\{\lambda_m\}$ of the polynomial are complex in \eqref{eqnresi}. 
For a pole of order $d$ \bblk{the probability grows polynomially as  $\Bar{P}_j \propto T^{2d-2}$. The growth is however much slower than the decay: still, $\bar{P}_j$ can exhibit sharp peaks due to these poles in the \textit{normalized vector} $\hat{\boldsymbol{\alpha}} (t)$ (see SI)}.

\textit{Perfectly ordered chains:}
Setting $s = -i\epsilon$ in Eq.~\eqref{eqnresi} yields the poles which correspond to crossing points as a function of $M(R_{jj'})$ (Fig.~\ref{fig:label1v1}a). In the nearest-neighbor (NN) regime, we find a \textit{single} crossing point, regardless of the atom number ($N=8$ is shown). 
By contrast, in the non-nearest neighbor (NNN) regime of long-range RDDI  (Fig.~\ref{fig:label1v1}b), there are multiple crossing points on account of the \textit{non-monotonic dependence} of $M (R_{jj'})$ on $k_0 R_{jj'}$. The corresponding time-averaged excitation probability of the initially excited atom has narrow resonances at inter-atomic separations $R_{jj'}$ or equivalently at RDDI values $M (R_{jj'})$ that correspond to those crossing points (Fig.~\ref{fig:prob_WOR}a). 

For long-ranged (NNN) RDDI, \textit{local minima} (anti-localization dips) are observed at the pseudo-crossings and local maxima  (localization) at the crossings of the time-averaged excitation probability (see Fig.~\ref{fig:prob_WOR}b) only if the dissipative effects are suppressed, as for $\omega_0$ inside a band gap~\cite{gogolin1982electron,abanin2019colloquium,vosk2015theory,kurizki_1990_pra}. Otherwise, large dissipation ($\Im M(k_0R)$) predominantly (but not completely) delocalizes the excitation (see below). 


\textit{Randomized chains:}
The changes in PIMATE due to position randomness are illustrated in Fig.~\ref{fig:pole_N=4_RR}.  The randomness transforms the crossing points in the eigenvalue spectrum into pseudo-crossing. Introducing randomness into the system causes eigenvalue repulsion~\cite{dyson1962statistical,10.1214/11-AOP710}. This repulsion eliminates any degeneracy in the system's eigenspectrum. Thus, a gap opens up at the crossing point when randomness is introduced in $M$ (Fig.~\ref{fig:pole_N=4_RR} a, b), transforming it into a pseudo-crossing point. The gap increases as randomness grows.  In the NNN regime of long-range RDDI, with randomized couplings, no crossing is observed; only multiple avoided crossings arise. Consequently, there are no distinct peaks in the time-averaged excitation probability distribution (Fig.~\ref{fig:pole_N=4_RR} c, d).

\textit{Dynamics:} The dynamics of the system can be described by a collection of snapshots of the excitation probability of atoms in the chain at different times at the crossover (resonance) point or non-resonance points and then time-averaged. The peaked recurrent excitation probability of the initially excited atom at the crossing point compared to an atom at a non-crossing point recorded at different time snapshots indicates partial trapping of the excitation at the crossover point (Fig.~\ref{Time_dynamics_M}).  

\bblk{The time-averaged excitation probabilities for $N=8$ show equivalent results for the time-dependent trapping in both the computational and the collective (Dicke) bases (see SI), consistently with Fig.~\ref{Time_dynamics_M}.  
}

\bblk{\subsection*{Fisher information on randomness}}
\bblk{The excitation probability of the initially excited TLA, $|\alpha_j(t)|^2$ can be used to \textit{estimate $M$ for the NN interaction} and $R_{ij}$ for the \textit{long-range NNN interaction} using the important tool of Fisher information~\cite{toth2014quantum,victor,Paris,PhysRevA.110.013715,pezze2018quantum} whereby, the precision of estimating $M$ or $R_{ij}$ is restricted by the Cram\'er-Rao bound on the error variance,
 $\text{Var}(M)\ge \frac{1}{\mathcal{N} H(M)}$,
where $H(M)$ being the relevant  Fisher information (FI) and $\mathcal{N}$ the number of measurements. Here we measure the time-averaged reduced density matrix $\bar{\rho}_j$ of the $j-$th atom, which is \textit{diagonal} in the computational (energy) basis. The FI for estimating $M$   can be expressed as~\cite{victor,braunstein1994statistical,correa2015individual}
\begin{subequations}
    \begin{equation}\label{Fisher}
 H_{M(R)}= \sum_{i=1}^2 \frac{|\frac{\partial  \bar{\rho}_j^{(i)}}{\partial M (R_{kk'})}|^2}{\bar{\rho}^{(i)}_j},
\end{equation}
where 
\begin{equation}
    \bar{\rho}_j^{(1)}=(\bar{\rho}_{ee})_j=|\alpha_j(t)|^2, \quad \rho_j^{(2)}=(\bar{\rho}_{rr})_j=1-|\alpha_j(t)|^2
\end{equation}
\end{subequations}
are the population of the j-th TLA state, averaged over a sufficiently long time-interval. }
\bblk{The motivation is to estimate the parameter $M(R_{ij})$ for the perfectly ordered and randomized chains with maximal precision.}

\bblk{The quantum Fisher information (QFI) is always greater than or equal to the classical Fisher information (CFI), by definition, as the QFI is maximized over all possible measurements of a parameter. If the CFI is strictly less than the QFI, it signifies that the measurement is not optimal for their parameter estimation due to quantum features.
Since here we measure state \textit{populations}, we need not distinguish between QFI and CFI, and aim for the maximal FI.}

\bblk{Let the position of the $j$~th atom be 
$X_j$, which randomly deviates from its mean position by $\delta X_j$ under the disorder.  The $\delta X_j$ are drawn from a Gaussian distribution around the mean $X_m$. We assume that $\delta X_j$ are much less than the mean separations, $R_{ij}$.  Such positional disorder results in a spread in the distribution of $M (R_{ij})$ in Eq. (2b) which is \textit{non-Gaussian} owing to the complicated dependence of $M(R_{ij})$ on $R_{ij}$  (although $R_{ij}=|X_i-X_j|$ are derived from a Gaussian distribution of the $\delta X_j$). This randomness in the chain is revealed by the FI, as shown here.}

 
For a specific atom-field coupling strength $\kappa$, the FI has a peak or divergence at a particular value of the RDDI $M$, which implies \textit{maximized precision of estimating} $M$. 
In the multi-atom chain, the degeneracies of the eigenvalues for certain values of the RDDI $M$ give rise to a high-order pole  in the residue and thus a peak in the time-averaged excitation probability (as shown in Fig.~\ref{fig:prob_WOR}) for the NN and long-range NNN interaction scenarios, both with and without randomness in the system. For this particular value of $M$, FI has a peak (Fig.~\ref{fig:fisher_N=4,8_WOR}) which occurs at the crossing points in the eigenvalue spectrum.  Therefore, by appropriately tuning the parameters of the system to the region where the FI peak occurs, the efficiency and precision of the excitation measurement are maximized. 



The washout of the excitation peak in Fig.~\ref{fig:pole_N=4_RR}c corresponds to the loss of FI divergence in the NN regime, which is a signature of randomness in the system. In the NNN long-range interaction regime, we observe diminished peaks of FI in the presence of randomness (Fig.~\ref{fig:fisher_N=4,8_WOR}).

The variation of the peak of the logarithmic FI with the randomness in $M$ for the NN interaction and with $R$ for the long-range NNN interaction is shown in the SI. With the increase in randomness, \textit{the peak value of the} FI \textit{ diminishes} (Fig.~\ref{fig:fisher_N=8_WVR}). This allows us to identify the presence of the amount of randomness in the network.


\section*{Discussion}

Precise control of interactions among atoms and their couplings to a cavity or waveguide fields is key to the success of quantum technology. For example, the fidelity of quantum gates in atomic quantum information processing and computation~\cite{zhang2022quantum,asadi2020protocols,saffman2016quantum,tiarks2019photon,su2023rabi} depends on the interatom resonant dipole-dipole interactions (RDDI). Hence, it is important to estimate the distribution of the RDDI and fine-tune it according to needs.

Towards these goals, we have revealed the hitherto unknown photon-induced many-atom trapped excitation (PIMATE) at the crossings of the field-dressed many-atom eigenvalues in the single-excitation domain (Fig.~\ref{fig:label1v1},\ref{fig:pole_N=4_RR}a,\ref{fig:pole_N=4_RR}b).  Retarded (long-ranged) RDDI among atoms has been shown to drastically affect PIMATE beyond the near zone (Fig.~\ref{fig:pole_N=4_RR}c,\ref{fig:pole_N=4_RR}d), Fig.~\ref{fig:prob_WOR}). 

It is important to distinguish PIMATE from \textit{static excitation trapping} at a particular site. As shown in the SI, in PIMATE the excitation oscillates in time (Fig.~\ref{Time_dynamics_M}), and trapping is only evident following long-time averaging.
Positional disorder in the multi-atom chain, which results in a spread in the RDDI; and the variability of the coupling strength of individual atoms to the field mode (SI) have been shown to be detrimental to PIMATE (Fig.~\ref{fig:pole_N=4_RR}-\ref{Time_dynamics_M}). Thus, \textit{randomization can be detected by the absence} of PIMATE, which provides a new perspective for distinguishing a localized phase from the ergodic phase~\cite{anderson_1958_pr,shepelyansky_1994_prl,abanin2019colloquium}.



Importantly, we have shown that PIMATE or its absence due to disorder can be monitored with enhanced precision or equivalently via very few (even single) measurements: the \textit{Fisher information (FI) diverges at a crossing of the photon-dressed energy eigenvalues} unless it is diminished bt disorder. 

\bblk{We stress that PIMATE does not conform to the accepted definitions of quantum phase transitions that occur, e.g. in the long-range interacting LMG model where a single crossing may occur~\cite{lipkin_1965, glick_1965,co_2018,pal_2023}. PIMATE exhibits multiple crossings, caused by the interplay between atom-field coupling and atom-atom RDDI, which has both short and long-range components.}

\bblk{To conclude, the concept of PIMATE introduced here can enhance our ability to sense and design interacting-atom arrays, by controlling the couplings of the atoms to a cavity~\cite{paternostro2009solitonic,ritter2012elementary} or waveguide~\cite{corzo2019waveguide,sheremet2023waveguide,shahmoon2016highly,shahmoon2011strongly} and/or their RDDI~\cite{PhysRevA.72.043803}. The proposed PIMATE is of interest since it constitutes a \textit{platform for a broad range of studies in quantum technologies} and the fundamentals of light-matter interaction, by highlighting many-body collective behavior in mesoscopic, $N \gg 1$ atom systems. PIMATE reveals new types of mesoscopic, collective dynamics that yield a much richer diversity of phenomena compared to that of few atoms, with both fundamental and technological consequences.}

Our findings are reminiscent of the role of\textit{ dynamical control} in enhancing the FI and thereby the precision of quantum thermometry \cite{mukherjee2019enhanced}, as well as the ability of \textit{criticality behavior} in quantum many-body systems to achieve enhanced-precision thermometry, magnetometry~\cite{coleman2005quantum,luca_2017_prl,ye_2020_phyicae}, or environment parameter estimation by a quantum probe~\cite{zwick2016maximizing,meher2024thermodynamic,PhysRevApplied.22.034058}.

\section*{Acknowledgment}
PC acknowledges support from the Ben May Center for Theory and Computation. DP would like to acknowledge the EU HORIZON-RIA project EuRyQa (grant No. 101070144) and DFG FOR 5413 project QUSP (grant No. 465199066). GK acknowledges support from DFG (FOR 2724).


%

\onecolumngrid
\appendix

\setcounter{equation}{0}
\newpage
\section*{Supplementary Information}
\renewcommand{\theequation}{S.\arabic{equation}}
\setcounter{figure}{0}
  \let\oldthefigure\thefigure
  \renewcommand{\thefigure}{S\oldthefigure}

\subsection*{Two-atom case}\label{appendix:A}
 Let us first consider the simplest scenario of only two TLAs. 
The Hamiltonian of the combined atom-cavity system\bb{~\cite{Gershon}} is then given by
 \begin{eqnarray} \nonumber
   H= \omega_1 \ketbra{e_1}{e_1}+ \omega_2\ketbra{e_2}{e_2} + M(\ketbra{e_1g_2}{g_1e_2}+H.c.)\\
 + \omega a_0^\dagger a_0 +\sum_\lambda(\kappa_{1\lambda}a_0\ketbra{e_1}{g_1}+
  \kappa_{2\lambda}a_0\ketbra{e_2}{g_2}+H.c.). \,\,\,
  \end{eqnarray}
  The Hamiltonian of the system leads to a general third-order polynomial equation for the evaluation of the eigenvalues, which of the form:
\begin{equation} \label{eq11111}
    \lambda^3- A \lambda^2 + B \lambda - C=0,
\end{equation}
$A = (\omega + \omega_1 +\omega_2)$, $B = (\omega_1+\omega_2) \omega + \omega_1 \omega_2- M^2- \kappa_A^2$, $C = (\frac{\omega_1+\omega_2}{2}) \omega + \kappa_A^2 (\frac{\omega_1+\omega_2}{2}) + 2 M\kappa_A \kappa_B - \kappa_B^2 - M \omega$. Here $\omega_1,\omega_2$ is the frequency of atom 1 and atom 2 respectively, and $\omega$ is the cavity frequency. To solve the general equation a variable change is executed, where we replace $\lambda = \Lambda + A/3$ and Eq.~\eqref{eq11111} reduces to 
\begin{equation}
    \Lambda^3 + p \Lambda + q = 0,
\end{equation}
where $p=(3B-A^2)/3$, and $q = (-2 A^3+ 9 A B -27 C)/27$. The roots are
\begin{eqnarray} \nonumber \label{eq1212}
    \Lambda_1 &= & \frac{1}{6^{2/3}(-9 q + \sqrt{12 p^3 +81 q^2})^{1/3}} \\ \nonumber &\times&  \Big[-2 \sqrt[3]{3} p + \sqrt[3]{2} \left(-9 q + \sqrt{12 p^3 +81 q^2} \right)^{1/3} \Big], \\ 
     \Lambda_2 & = & \frac{1}{6^{2/3}(-9 q + \sqrt{12 p^3 +81 q^2})^{1/3}} \\ \nonumber & \times & \Big[ \sqrt[3]{-1} \left(2 \sqrt[3]{3} p + \sqrt[3]{-2} (-9 q + \sqrt{12 p^3 +81 q^2})^{1/3} \right)   \Big],\\ \nonumber
     \Lambda_3 &= & \frac{1}{6^{2/3}(-9 q + \sqrt{12 p^3 +81 q^2})^{1/3}}\\ \nonumber
     & \times & \Big[ \sqrt[3]{-1} \left(2 \sqrt[3]{-3} p + \sqrt[3]{2} (-9 q + \sqrt{12 p^3 +81 q^2})^{1/3} \right)   \Big]
\end{eqnarray}
For the specific values of the parameters as shown in Fig.~\ref{fig:label1}c, we get a crossing, i.e., a degeneracy in the eigenvalues from Eq.~\eqref{eq1212}. For the far zone case ($\kappa_a \neq \kappa_B$) there is no crossing in the eigenvalues.

{\it In Laplace domain:} The solution of the combined field-atom wave function $\ket{\psi(t)}$ in the Laplace domain can be written as in Eq.~\eqref{eqn22}.
 \begin{eqnarray}
 \vert \psi(t) \rangle = \sum^2_{j=1} \alpha_j (t) \vert r_j\rangle + \beta (t) \vert 1\rangle.     
 \end{eqnarray}
Using Schrodinger's equation for time evolution we can rewrite the time-dependent coefficients $\alpha_j$ and $\beta$ in the following form
\begin{eqnarray}
 &&\dot{\alpha}_j = -i \omega_0 \alpha_j - i\kappa_j\beta - i \sum_{j'} M_{jj'} \alpha_{j'}, {~~} \nonumber\\
 &&\dot{\beta} = -i \beta \omega - i \sum_j \alpha_j \kappa^*_j. \quad j=1,2
\end{eqnarray}
Further, in the Laplace domain\bb{~\cite{kofman1994spontaneous}}, the above set of coupled differential equations transforms into a set of coupled algebraic equations of the form, 

\begin{eqnarray}
 &&s\hat{\beta} (s) = -i \omega \hat{\beta}(s) - i \sum_j \kappa^*_j \hat{\alpha}_j(s)  + \beta(0),  j=1,2 \nonumber\\
&&  s \hat{\alpha}_j(s) = -i \omega_0 \hat{\alpha}_j(s) - i\kappa_j \hat{\beta}(s) - i \sum_{j'} M_{jj'} \hat{\alpha}_{j'}(s) + \alpha_j(0). \nonumber\\
\end{eqnarray}
 
The solution to $\hat{\boldsymbol{\alpha}}(s)$ in the two-atom case is
\begin{eqnarray}\label{eqa111}
&&\hat{\alpha}_1(s) = \vert \mathcal{A}(s)\vert^{-1} (s+i\omega_2 + \frac{|\kappa_2|^2}{s+i\omega}), \nonumber\\
&&\hat{\alpha}_2(s) = -\vert \mathcal{A}(s)\vert^{-1},\nonumber\\
&& |\mathcal{A}(s)| = M^2 - \frac{iM(\kappa^*_1\kappa_2+\kappa^*_2\kappa_1)}{s+i\omega}+(s+i\omega_1)(s+i\omega_2) \nonumber \\
&&+ \frac{|\kappa_2|^2(s+i\omega_1)}{s+i\omega} + \frac{|\kappa_1|^2(s+i\omega_2)}{s+i\omega}.
\end{eqnarray}

Setting $s = -i\lambda$ in $|\mathcal{A}(s)| = 0$, gives the poles as a function of $M$.
In the above expression, the polynomial has \textit{both linear and oscillatory terms for simple poles}. If there is at least one pole of order two in $\vert \mathcal{A}(s) \vert$ the polynomial is cubic rather than linear, apart from the oscillatory terms. The degeneracy of the eigenvalues occurs at a particular value of $M$, say $M_{cr}$ the crossing point of the eigenvalues.  

\subsection*{ Field-dressed multiatom network resonance dynamics via inverse Laplace transform}\label{appendix:B}


 

The solution of  $\hat{\alpha} (s)$ for $N$ atom system is 
\begin{eqnarray}
 \hat{\boldsymbol{\alpha}}(s) =  \frac{\rm{adj}(A(s))\boldsymbol{\alpha}(0)}{\vert A(s) \vert} 
\end{eqnarray}

The above ratio of the two polynomials is a proper fraction since the degree of the denominator polynomial exceeds that of the numerator: the numerator and the denominator polynomials have maximum degrees $ s^{2n-2}$ and $s^{2n}$ respectively. 

If $\lambda_{m}$ is the $m$~th pole of order $k_m$ of the denominator polynomial, which has $l$ distinct poles, the fraction can be expanded in the following partial fraction form,

\begin{eqnarray}
 \hat{\boldsymbol{\alpha}}(s) =   \frac{\rm{adj}(A(s))\boldsymbol{\alpha}(0)}{\prod^l_{m=1}(s-\lambda_{m})^{k_m}} =  \sum^l_{m=1} \sum^{k_m}_{j=1} \frac{\boldsymbol{r}_{m, j}}{(s-\lambda_{m})^{j}},
\end{eqnarray}

where, the $k_q$ residues for $q$~th pole are computed using the following formula

\begin{eqnarray}
 \boldsymbol{r}_{q, j} = \frac{1}{(k_q-j)!} \frac{d^{k_q-j}}{ds^{k_q-j}}\left(  \frac{\rm{adj}(A(s))\boldsymbol{\alpha}(0)}{\prod^l_{m=1; m\ne q}(s-\lambda_{m})^{k_m}}\right) \Bigg|_{s=\lambda_q}, {~~~~} j = 1,...,k_q.   
\end{eqnarray}
The inverse Laplace transform of the above expression yields,

\begin{eqnarray}
 \boldsymbol{\alpha}(t) = \sum^l_{m=1} \sum^{k_m}_{n=1} \boldsymbol{r}_{m,n}  \mathcal{L}^{-1} \left[ \frac{1}{(s-\lambda_{m})^{n}}  \right]=  \sum^l_{m=1} \sum^{k_m}_{n=1} \frac{\boldsymbol{r}_{m,n}}{(j-1)!} t^{n-1}e^{\lambda_m t} = \sum^l_{m=1} \sum^{k_m}_{j=1} \boldsymbol{A}_{m,n} t^{n-1}e^{\lambda_m t}. 
\end{eqnarray}

Therefore, the transition probability of the first atom can be written as

\begin{eqnarray}
  P_j= \int^T_0 \vert  \alpha_j(t) \vert^2 dt =  \sum^l_{m=1} \sum^l_{m'=1} \sum^{k_m}_{n=1}  \sum^{k_{m'}}_{n'=1} {A^{1*}}_{m',n'} {A^1}_{m,n} \int^T_0 dt~t^{(n+n')-2}e^{t(\lambda^*_{m'} +\lambda_m)}.
\end{eqnarray}

Solving the definite integral,  we obtain
\begin{subequations}
\begin{eqnarray}
 P_j=  \int^T_0 \vert  \alpha_j(t) \vert^2 dt = \sum^l_{m=1} \sum^l_{m'=1} \sum^{k_m}_{n=1}  \sum^{k_{m'}}_{n'=1} {A^{1*}}_{m',n'} {A^1}_{m,n} I_{m,m',n,n'}(T),\nonumber\\
\end{eqnarray}

where~\footnote{Note: $\int t^n e^{at} dt = e^{at} \sum^n_{p=0} t^p (-1)^{n-p} \frac{n!}{p! a^{n-p+1}}$ for $n>0, a \ne 0$}
\begin{eqnarray}
    I_{m,m',n,n'}(T) & = & \sum^{(n+n')-2}_{p=0}\frac{(-1)^{(n+n')-p-2} (n+n'-2)!}{p!((\lambda^*_{m'}+\lambda_{m}))^{(n+n')-p-1}} e^{T(\lambda^*_{m'} + \lambda_m)} T^p \mbox{~~~for~~~}  n+n' > 2, \lambda_m + \lambda^*_{m'} \ne 0, \nonumber\\
    I_{m,m',n,n'}(T) & = & \frac{e^{T(\lambda^*_{m'} + \lambda_m)} - 1}{(\lambda^*_{m'} + \lambda_m)} \mbox{~~~for~~~}  n+n' = 2, \lambda_m + \lambda^*_{m'} \ne 0,  \\
    I_{m,m',n,n'}(T) & = & \frac{T^{n+n'-1}}{n+n'-1}\mbox{~~~for~~~} \lambda_m + \lambda^*_{m'} = 0. \nonumber
\end{eqnarray}
\end{subequations}
This crossing in turn gives rise to divergence in the residue. Equation (14b) averaged over time $T$ will then be given by  
\begin{eqnarray}
  && \bar{P}_j= \frac{1}{T} \int^T_0 \vert  \alpha_j(t) \vert^2 dt = \sum^l_{m=1} \sum^l_{m'=1}\sum^{k_m}_{n=1} \sum^{k_{m'}}_{n'=1} {A^{1*}}_{m',n'} {A^1}_{m,n} \frac{T^{n+n'-2}}{n+n'-1} \nonumber\\ &&+ \sum^l_{m=1} \sum^l_{m'=1} \sum^{k_m}_{n=1} \sum^{k_{m'}}_{n'=1} \sum^{n+n'-2}_{p=0} {A^{1*}}_{m',n'} {A^1}_{m,n} \frac{(-1)^{(n+n')-p-2} (n+n'-2)!}{p!((\lambda^*_{m'}+\lambda_{m}))^{(n+n')-p-1}} T^{p-1} e^{T(\lambda^*_{m'} + \lambda_{m})} \nonumber\\
  && + \frac{1}{T} \sum_{m,m'}  {A^{1*}}_{m',n'} {A^1}_{m,n} \frac{e^{T(\lambda^*_{m'} + \lambda_{m})}}{\lambda_m + \lambda^*_{m'}} + \mbox{constant},\nonumber\\
\end{eqnarray}

which can be rewritten as
\begin{eqnarray}
  &&\bar{P}_j = \frac{1}{T} \int^T_0 \vert  \alpha_j(t) \vert^2 dt = \sum^{k_m}_{n=1} \sum^{k_{m'}}_{n'=1} X_{n,n'} T^{n+n'-2} \nonumber\\ &&+ \sum^l_{m=1} \sum^l_{m'=1} \sum^{k_m}_{n=1} \sum^{k_{m'}}_{n'=1} \sum^{n+n'-2}_{p=0} Y_{m,m',n,n',p}T^{p-1} e^{T(\lambda^*_{m'} + \lambda_{m})} \nonumber\\
  && + \frac{1}{T} \sum_{m,m'} Z_{m,m'} e^{T(\lambda^*_{m'} + \lambda_{m})} + \mbox{constant},\nonumber\\
\end{eqnarray}
\bblk{where $X_{nn'}$, $Y_{mm',nn',p}$, and $Z_{mm'}$ are the elements of real, positive matrices. The terms containing $e^{T(\lambda^*_{m'} + \lambda_{m})}$ factors are oscillatory since $\text{Im} (\lambda^*_{m'} + \lambda_{m}) \neq 0$ but also decaying for $\text{Re} (\lambda^*_{m'} + \lambda_{m}) < 0$, which are the only physical solutions. There is polynomial growth due to $T^{n+n'-2}$ and $T^{p-1}$ for $n+n'>2$ and $p>1$ respectively. }

\subsection*{Effect of Randomness}

The effects of randomness in PIMATE are shown in Fig.~\ref{fig:pole_N=4_RR}. This randomness transforms the crossing points in the eigenvalue spectrum into pseudo-crossings. As a result, distinct peaks in the time-averaged excitation probability distribution are no longer observed.

\bblk{\subsection*{Non-dissipative effect}
\begin{eqnarray}\label{EqM}
   && \text{Re} M_{ij} (\theta,R_{ij}) = J_{ij}(\theta,R_{ij}),
\end{eqnarray}
where 
\begin{eqnarray}
 J_{ij}(\theta,R_{ij}) &=& \frac{\gamma}{4} \Bigg[(1-3\cos^2\theta) \left(\frac{\cos (k_0 R_{ij})}{(k_0 R_{ij})^3} + \frac{\sin(k_0 R_{ij})}{(k_0 R_{ij})^2}\right) \nonumber\\
  && - (1-  \cos^2\theta) \frac{\cos (k_0 R_{ij})}{(k_0 R_{ij} )} \Bigg];
  \end{eqnarray}
  where $\gamma$ is the single-atom radiative decay rate at $\omega_0$, and $\theta$ is the angle between the direction of atomic dipole moment and the interatomic axis.}

\subsection*{Dissipative effects}
\begin{figure*}
\centering
         \subfigure[]{%
  \includegraphics[width=0.48\textwidth, height=0.235\textheight]{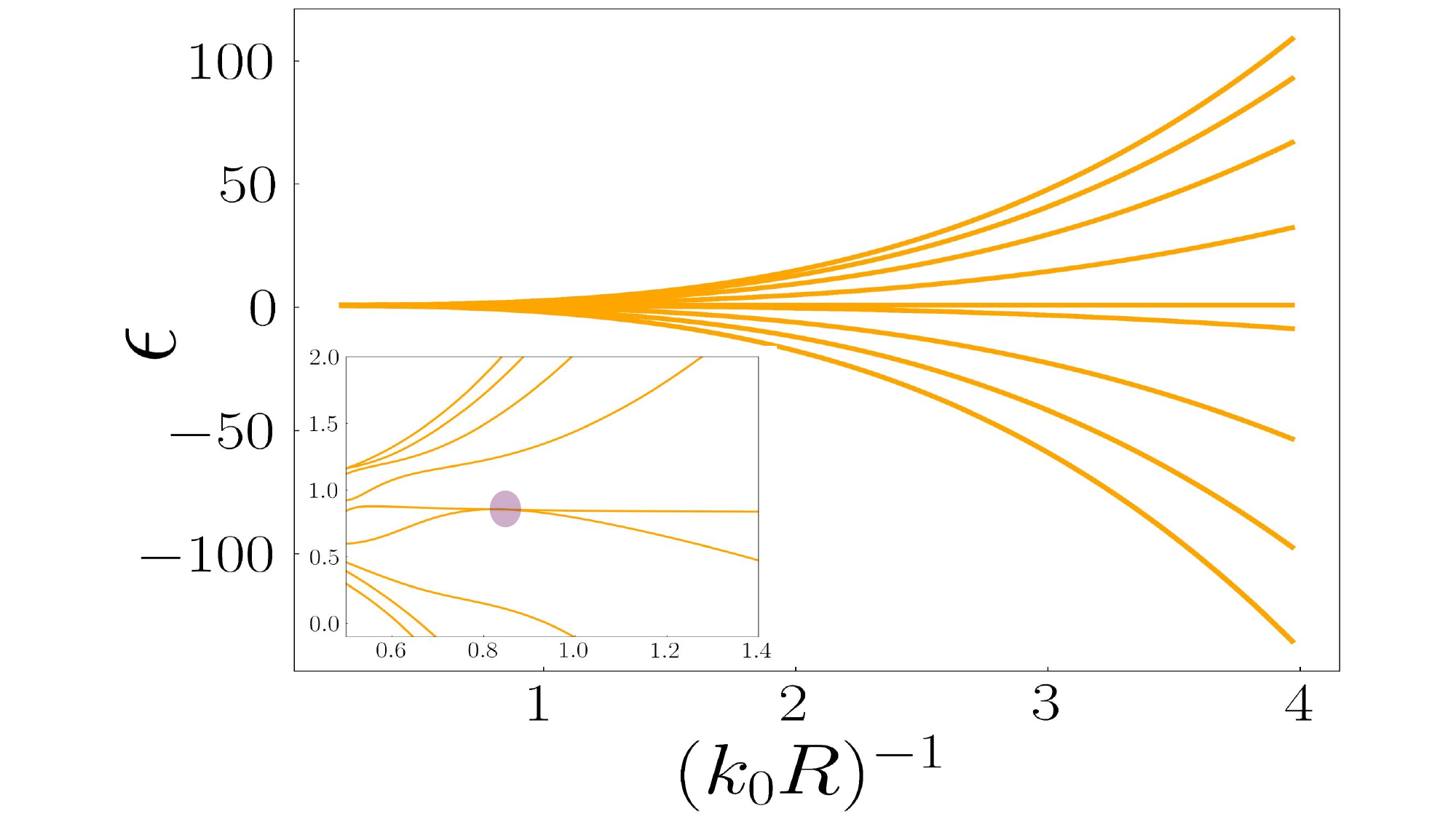}
  }
 \subfigure[]{%
  \includegraphics[width=0.4\textwidth]{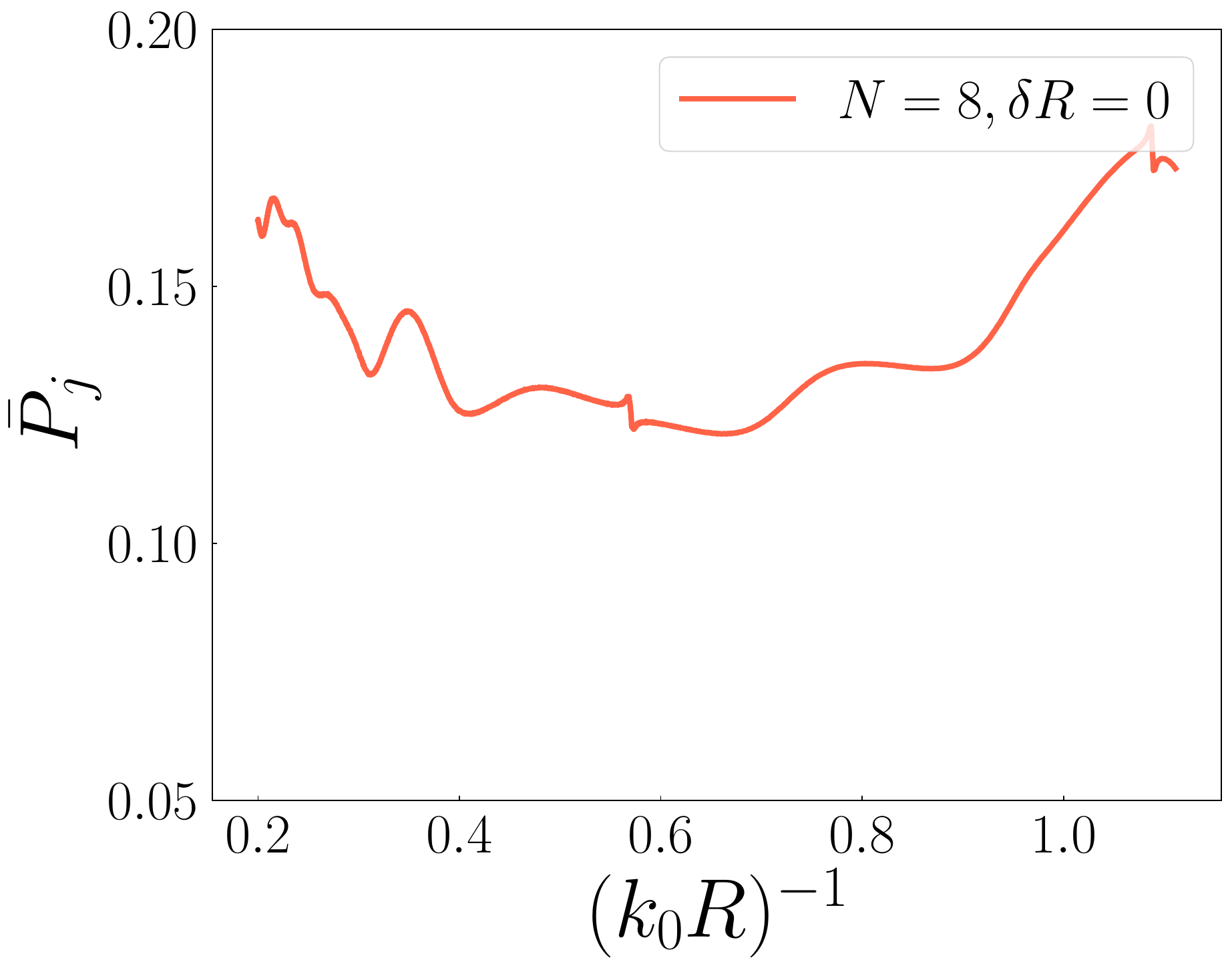}
  }
  \subfigure[]{%
  \includegraphics[width=0.45\textwidth]{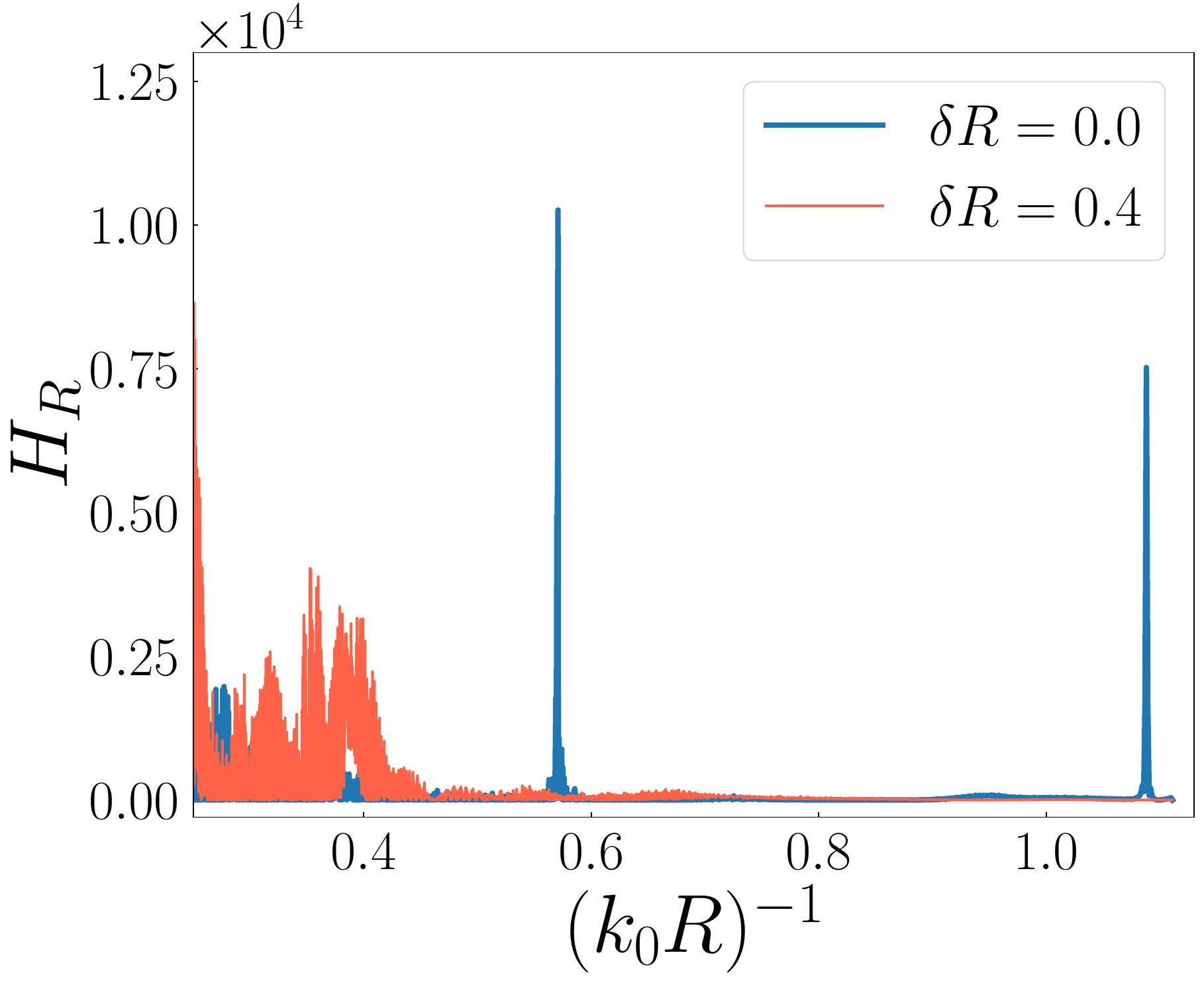}}
\caption{(a) Change of crossing points to pseudo-crossings as a function of separation $k_0 R$ \bblk{(in units of $\gamma$)}  for large dissipation (collective dissipation rate  $\gamma_{ij}/M \rightarrow 1$). The pseudo-crossing points of eigenvalues on all plots are marked by circles. (b) Resonances of the time-averaged excitation probability as a function of $k_0 R$ for large dissipation (collective dissipation rate $\gamma_{ij}/M \rightarrow 1$). The dips exhibit suppressed  \textit{(anti-) trapping}. (c) Variation of the quantum Fisher information for large dissipation ($\gamma_{ij}/M \rightarrow 1$) in the presence of off-diagonal randomness (red solid line) and without randomness (blue solid line).}
\label{Dissipation_part}
\end{figure*}

In the long-range NNN regime, if we consider the collective dissipation to be large, $\gamma \sim \gamma_{ij} \sim M_{ij}$, then the crossing points disappear due to the decay, but pseudo crossing points are still observed for $N=8$ for the perfectly ordered chains (Fig.~\ref{Dissipation_part}a). In the case of the randomized chains, even the pseudo-crossing points get washed out. Namely, randomness along with dissipation washout the effect of pseudo-crossing from the eigenvalue spectrum. Excitation delocalization due to the presence of large dissipation in the long-range NNN interaction regime, without the randomness is shown in Fig.~\ref{Dissipation_part}b. For polarization parallel or perpendicular to the inter-atom axis the matrix of radiative dissipative rates is~\cite{lehmberg_pra_1970} 

\begin{subequations}
\begin{eqnarray}
 \gamma_{ij} = \gamma f_{\Sigma} (k_0R_{ij}),\nonumber\\
 \gamma_{ij} = \gamma f_{\Pi} (k_0R_{ij}),
\end{eqnarray}
 where
\begin{eqnarray}  \nonumber
&& f_{\Sigma} (k_0 R_{ij}) = 3 \left[ \frac{\sin(k_0R_{ij})}{(k_0R_{ij})^3} - \frac{\cos(k_0R_{ij})}{(k_0R_{ij})^2}\right], \\
&& f_{\Pi} (k_0 R_{ij}) = -\frac{3}{2}\left[\frac{\sin (k_0 R_{ij})}{(k_0 R_{ij})^3} - \frac{\cos(k_0 R_{ij})}{(k_0 R_{ij})^2} - \frac{\sin (k_0 R_{ij})}{(k_0 R_{ij} )} \right].
\end{eqnarray}
\end{subequations}
In the near zone,  $k_0R_{ij}<<1$, the radiative rates $\gamma_{ij}/\gamma = f_{\Pi (\Sigma)} (k_0 R_{ij}) \rightarrow 1$ are \textit{negligible} compared to the RDDI $\text{Re}(M_{ij}) \sim \cos (k_0 R_{ij})/(k_0 R_{ij})^3$. In the far zone, $k_0R_{ij} \gtrsim 1$, the ratio $\gamma_{ij}/M_{ij} \rightarrow 1$ and we may estimate the dissipative effects by replacing $\gamma_{ij} \rightarrow \gamma$. However, $\gamma$ and $\gamma_{ij}$ can be drastically \textit{suppressed} in cavities or waveguides by placing the resonant transition frequency $\omega_0 = k_0 c$ inside a \textit{photonic band gap} where $M_{ij}$ remains essentially undiminished~\cite{shahmoon2011strongly, shahmoon2016highly,kofman_kurizki, kurizki_1990_pra}. 

Thus, we can address one of two possible regimes: either negligible dissipation compared to RDDI, $\gamma_{ij}<< M_{ij}$, or dissipation that is comparable to RDDI, $\gamma \gtrsim \gamma_{ij} \sim M_{ij}$. The effects of dissipation in PIMATE are shown in Fig.~\ref{Dissipation_part}. In the presence of dissipation transforms the crossing points in the eigenvalue spectrum into pseudo-crossings. As a result, distinct peaks in the time-averaged excitation probability distribution are no longer
observed. With the randomness being introduced along with the presence of dissipation we don't have even pseudo-crossing thus a loss of QFI divergence.




\vspace{10ex}

\subsection*{Time-domain evolution in the computational and Dicke bases}
The atom-field system with atom 1 initially excited, the state vector for the atom-field system at
time $t$~\cite{scully2007correlated,svidzinsky2008dynamical} has the following state in the Dicke basis
\begin{eqnarray} 
   \vert \psi (t)\rangle &= & \left[\eta_{+}(t) \vert +\rangle + \sum_{j=1}^{N-1} \zeta_j \vert j\rangle \right] \vert 0\rangle +\nu (t) \vert g,\cdots,g\rangle \vert 1 \rangle.
\end{eqnarray}
Here first term with amplitude $\eta_+ (t)$ corresponds to the state which is prepared by uniformly absorbing a single photon with wavevector $k_0$, $\vert +_{k_0}\rangle = \frac{1}{\sqrt{N}} \sum_k e^{ik_0 X_k} \vert e,\cdots,r_k,\cdots, e\rangle$. The subsequent terms with amplitude $\zeta_j$ corresponds to Dicke states $\j\rangle$, where $\vert j=1\rangle = \frac{1}{\sqrt{2}} \left[e^{ik_0 X_1} \vert r_1\rangle -e^{ik_0 X_2} \vert r_1\rangle \right]$  and the state $\vert j= N-1\rangle = \frac{1}{\sqrt{N-1}} \Big[e^{ik_0 X_1}|r_1\rangle + e^{ik_0 X_2}|r_2\rangle + e^{ik_0 X_3}|r_3\rangle +\cdots + e^{ik_0 X_{N-1}}|r_{N-1}\rangle - (N-1) e^{ik_0 X_N}|r_{N}\rangle \Big]$.
\begin{figure*}[ht]
  \begin{center}
    \subfigure[]{%
  \includegraphics[width=0.49\textwidth]{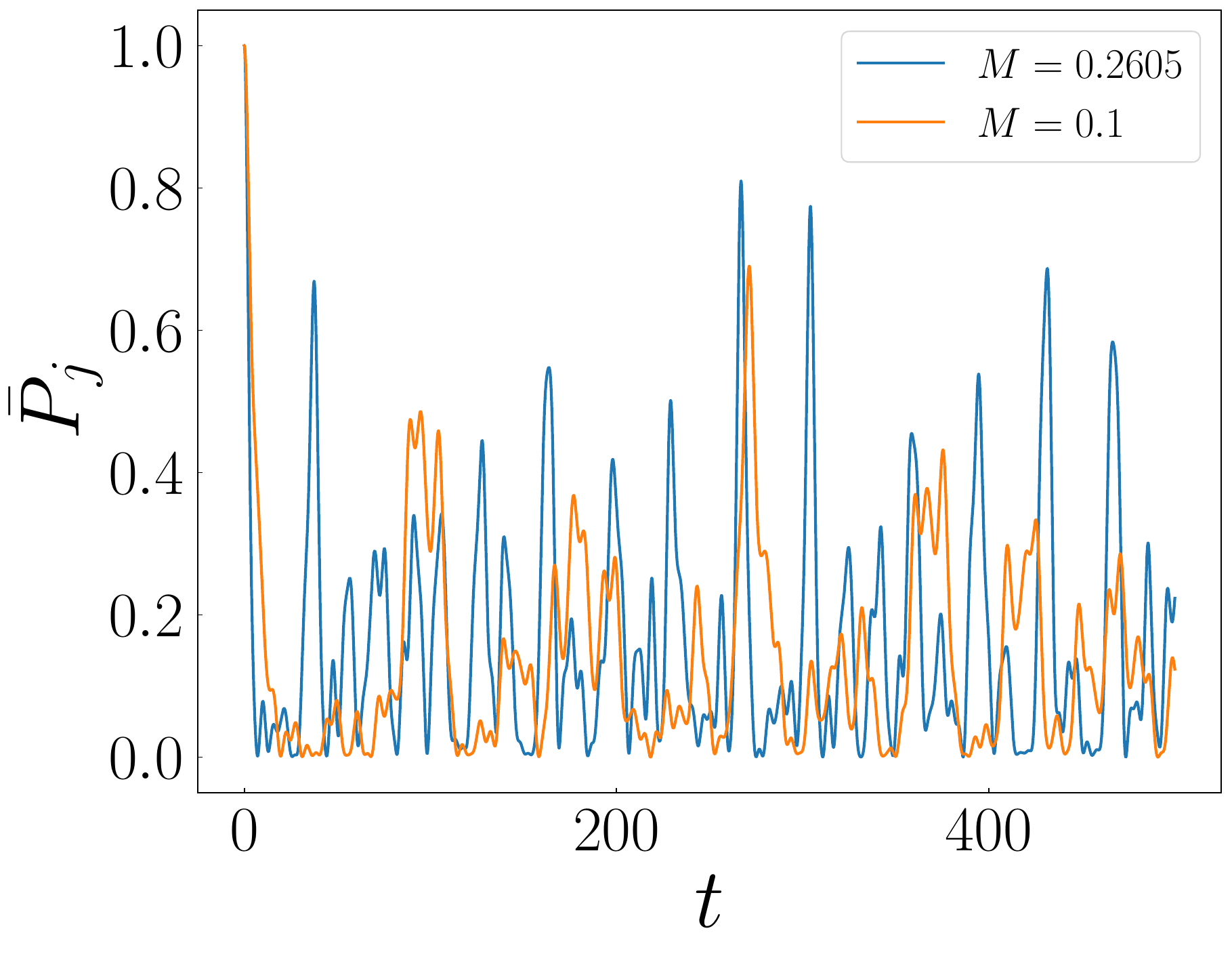}
  }
  \subfigure[]{%
  \includegraphics[width=0.49\textwidth]{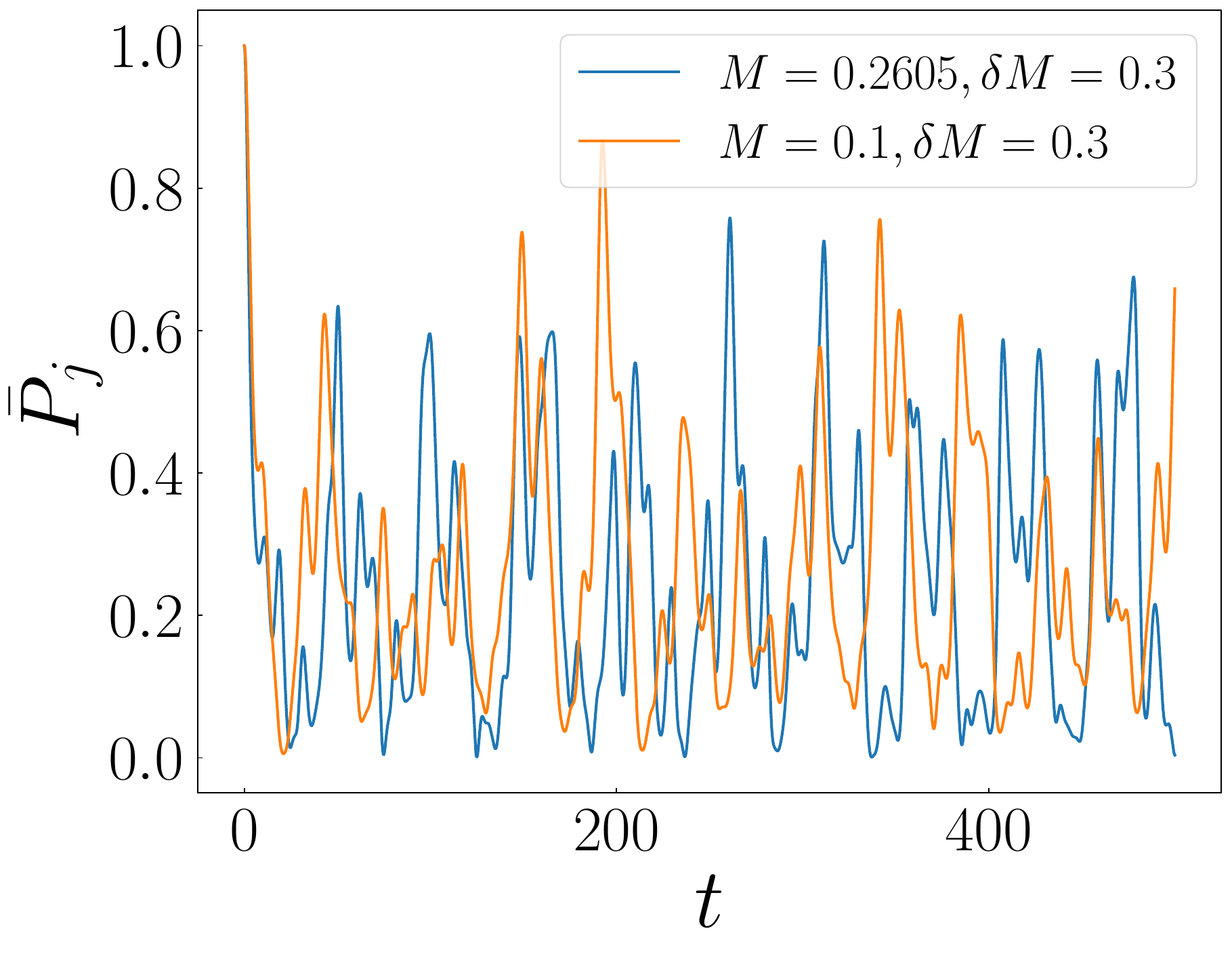}
  }
\end{center}
      \caption{(a) Time dependence of excitation probability for the first atom out of $N=4$ at the crossing point (resonance point) $M=M_{cr}$ (blue solid line) and non-resonance point (orange solid line). (b) Same in the presence of randomness at the crossing point (blue solid line) and non-resonance point (orange solid line). The parameter are as in Fig.~\ref{Time_dynamics_M}.}
      \label{fig:probWtime_wt}
  \end{figure*}

\begin{figure*}
  \begin{center}
    \subfigure{%
  \includegraphics[scale=0.195]{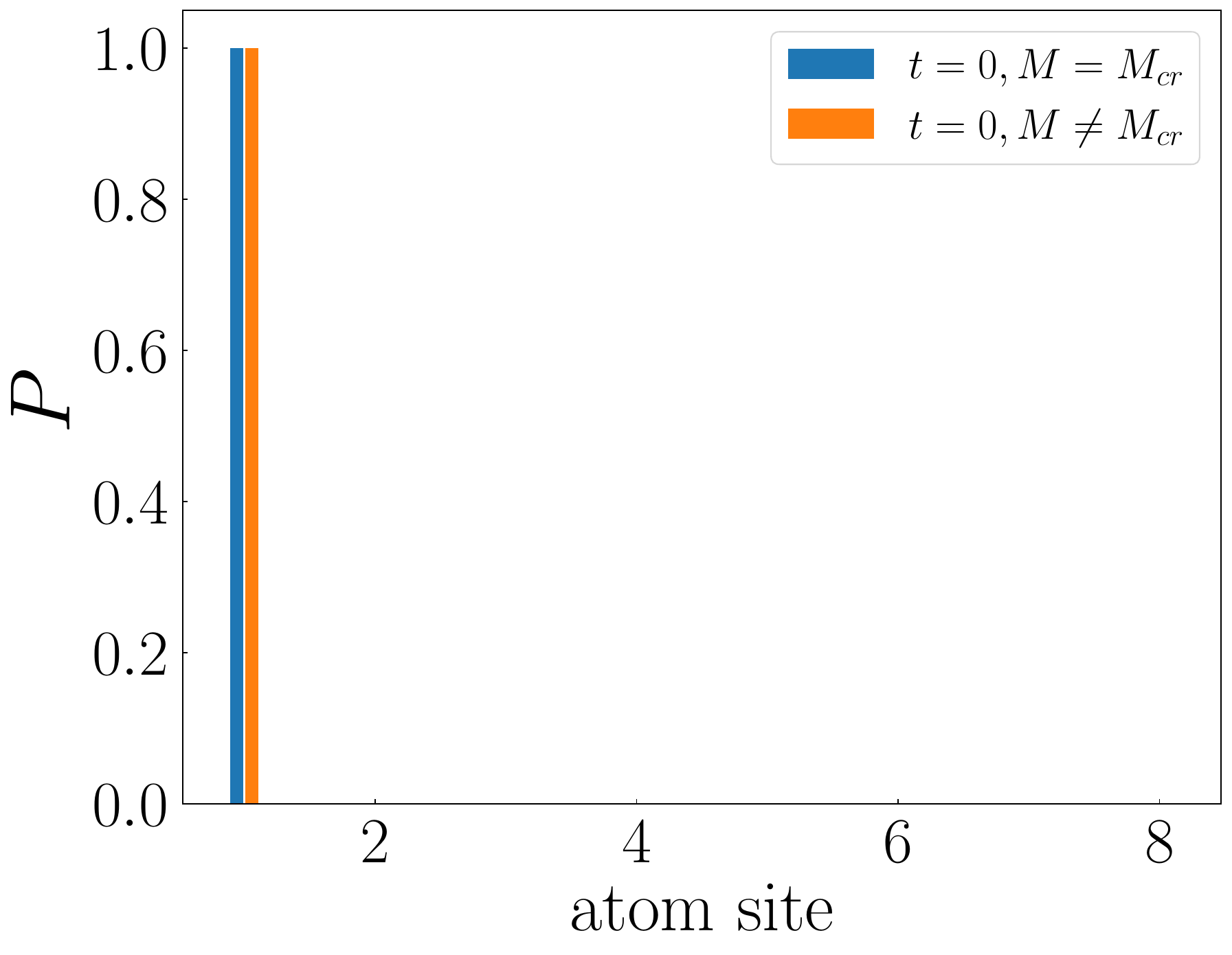}}
  \subfigure{%
  \includegraphics[scale=0.195]{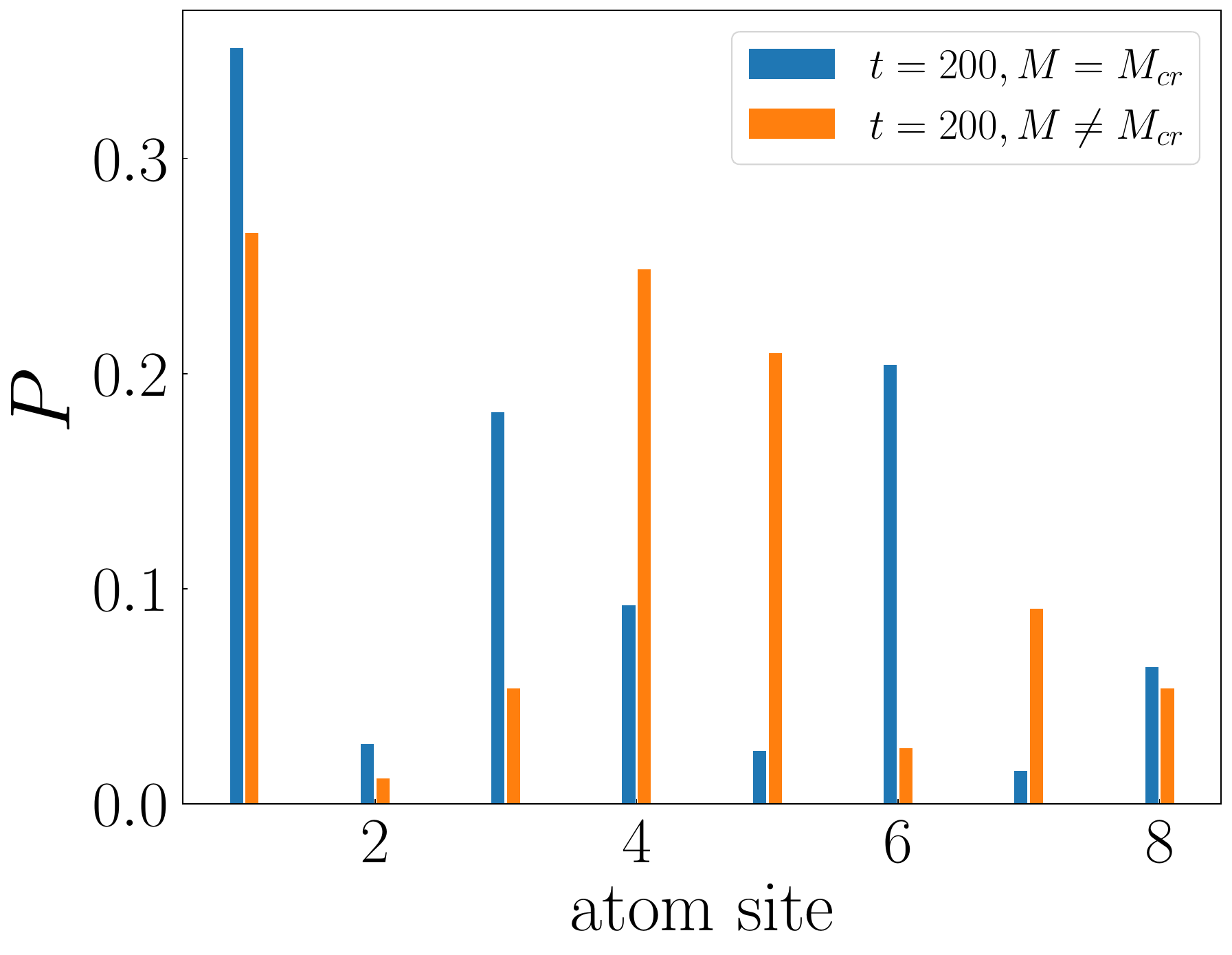}}
  \subfigure{%
  \includegraphics[scale=0.195]{prob_M_merged_t=1000.pdf}}\\
  \subfigure{%
  \includegraphics[scale=0.195]{prob_M_merged_t=100.pdf}}
   \subfigure{%
  \includegraphics[scale=0.195]{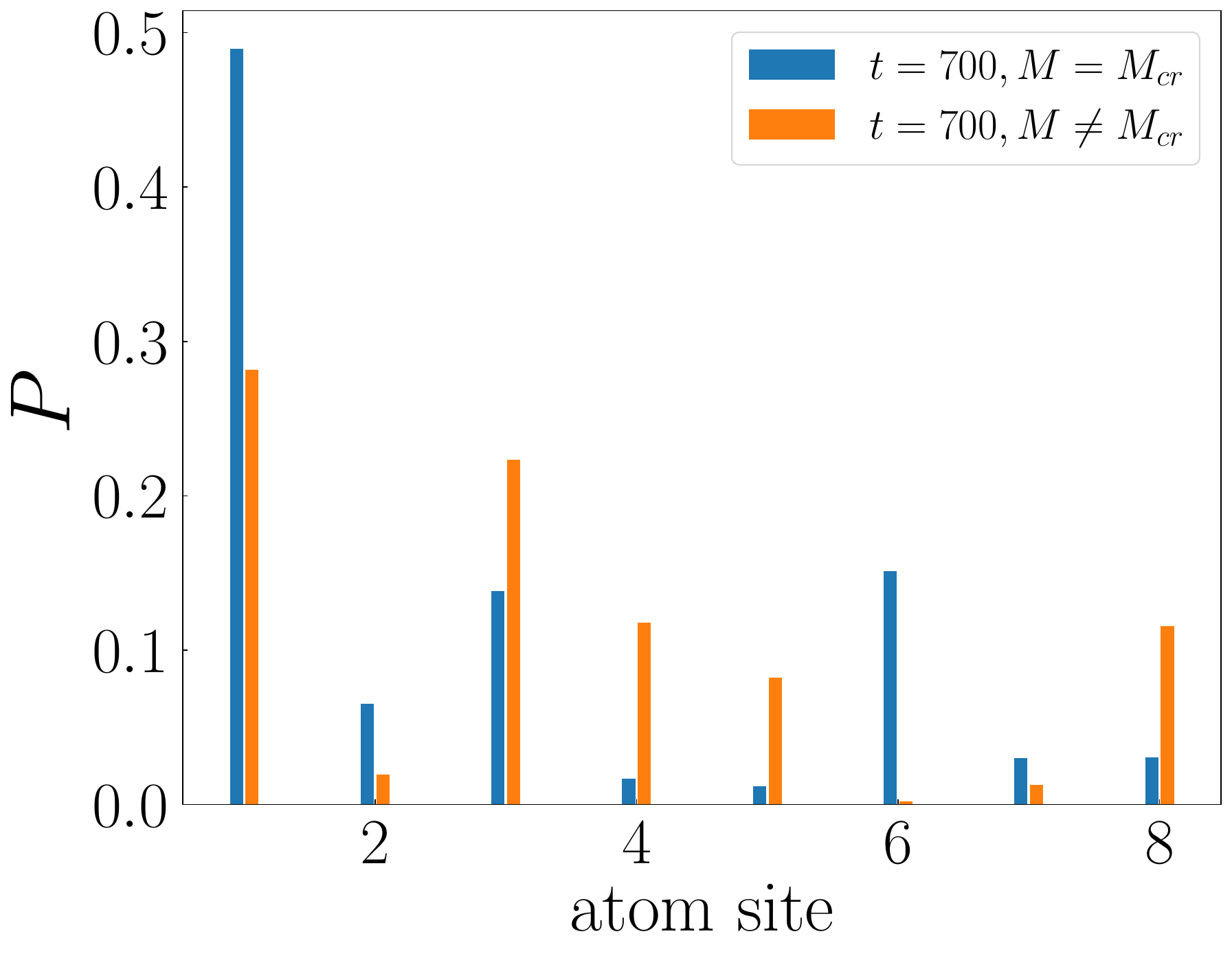}}
  \subfigure{%
  \includegraphics[scale=0.195]{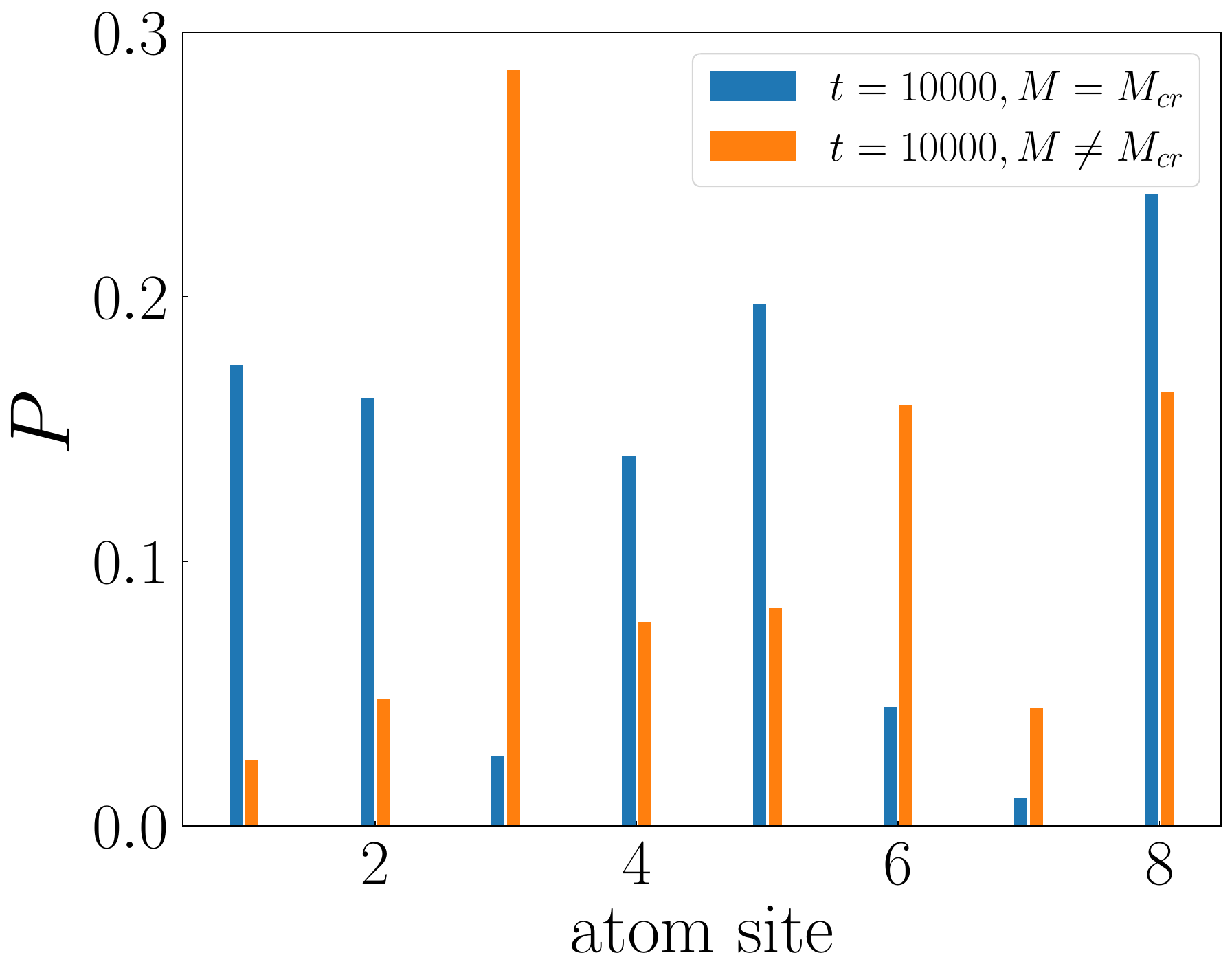}}
\end{center}
      \caption{ Time dependence of excitation probability with the atom sites at the crossing point $M=M_{cr}$ in blue. Same at a non-resonance point in orange. Fig.~\ref{Time_dynamics_M}.}
      \label{fig:probWtime_wt}
  \end{figure*}

\begin{figure*}
  \begin{center}
  \subfigure{%
  \includegraphics[scale=0.195]{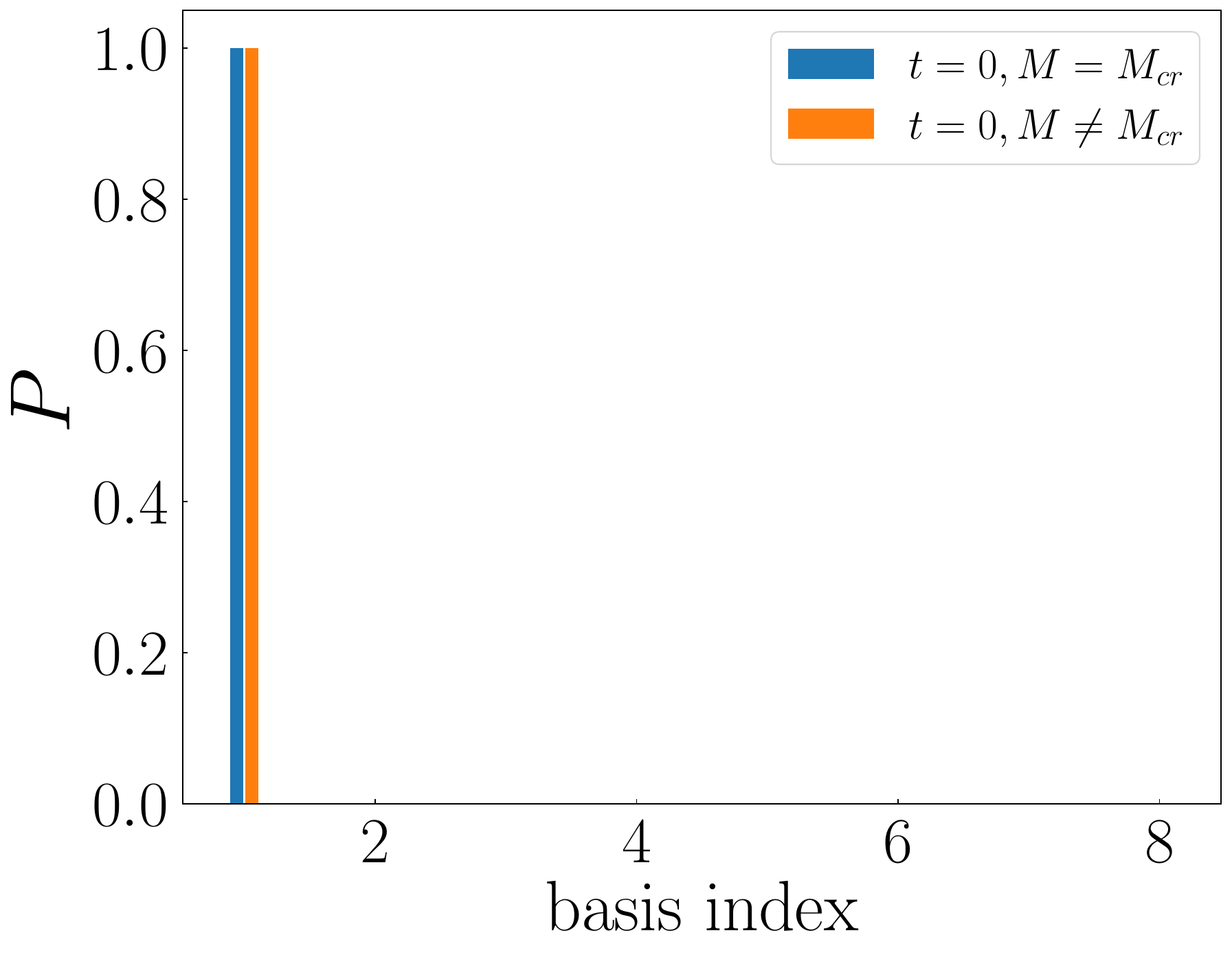}}
  \subfigure{%
  \includegraphics[scale=0.195]{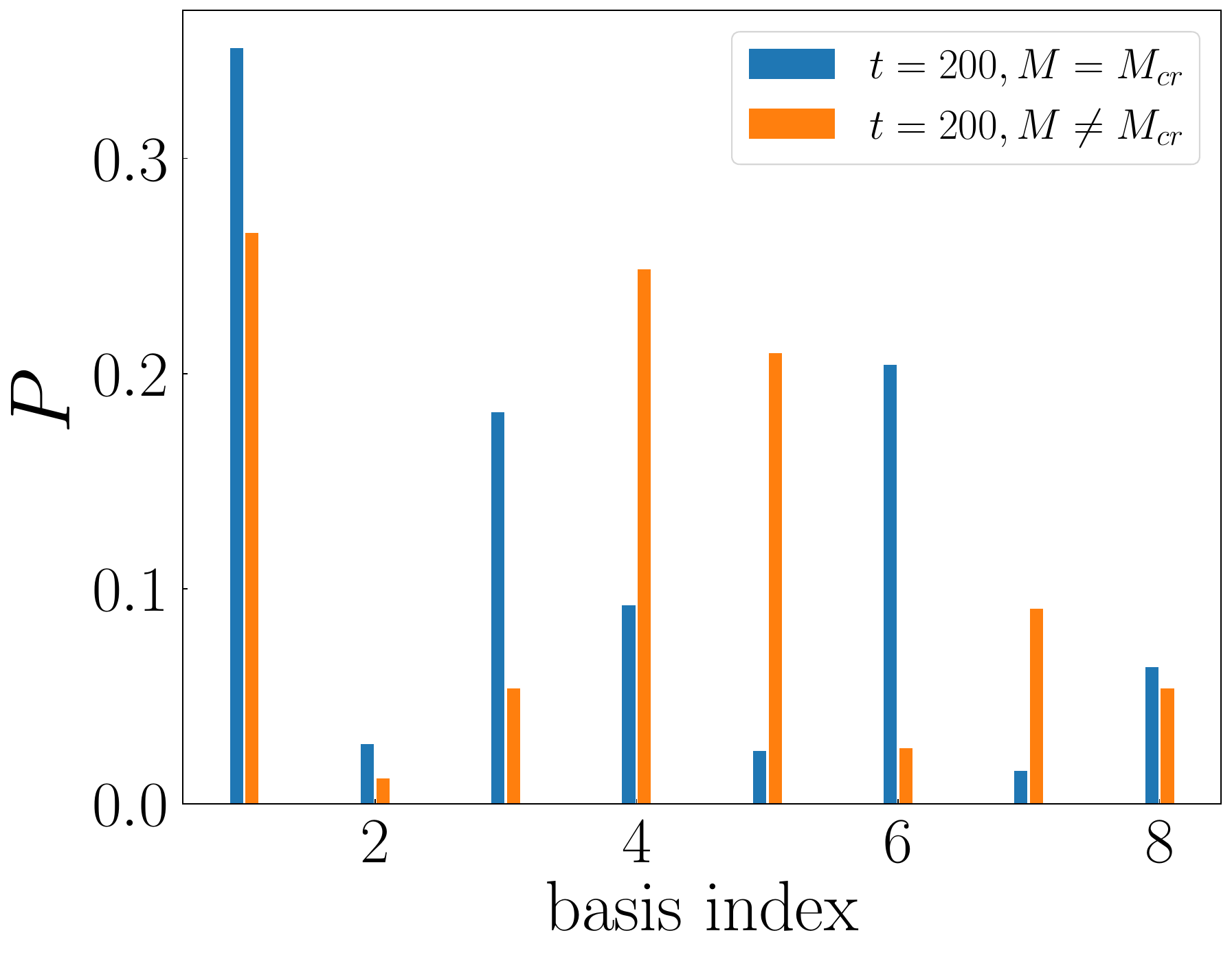}}
   \subfigure{%
   \includegraphics[scale=0.195]{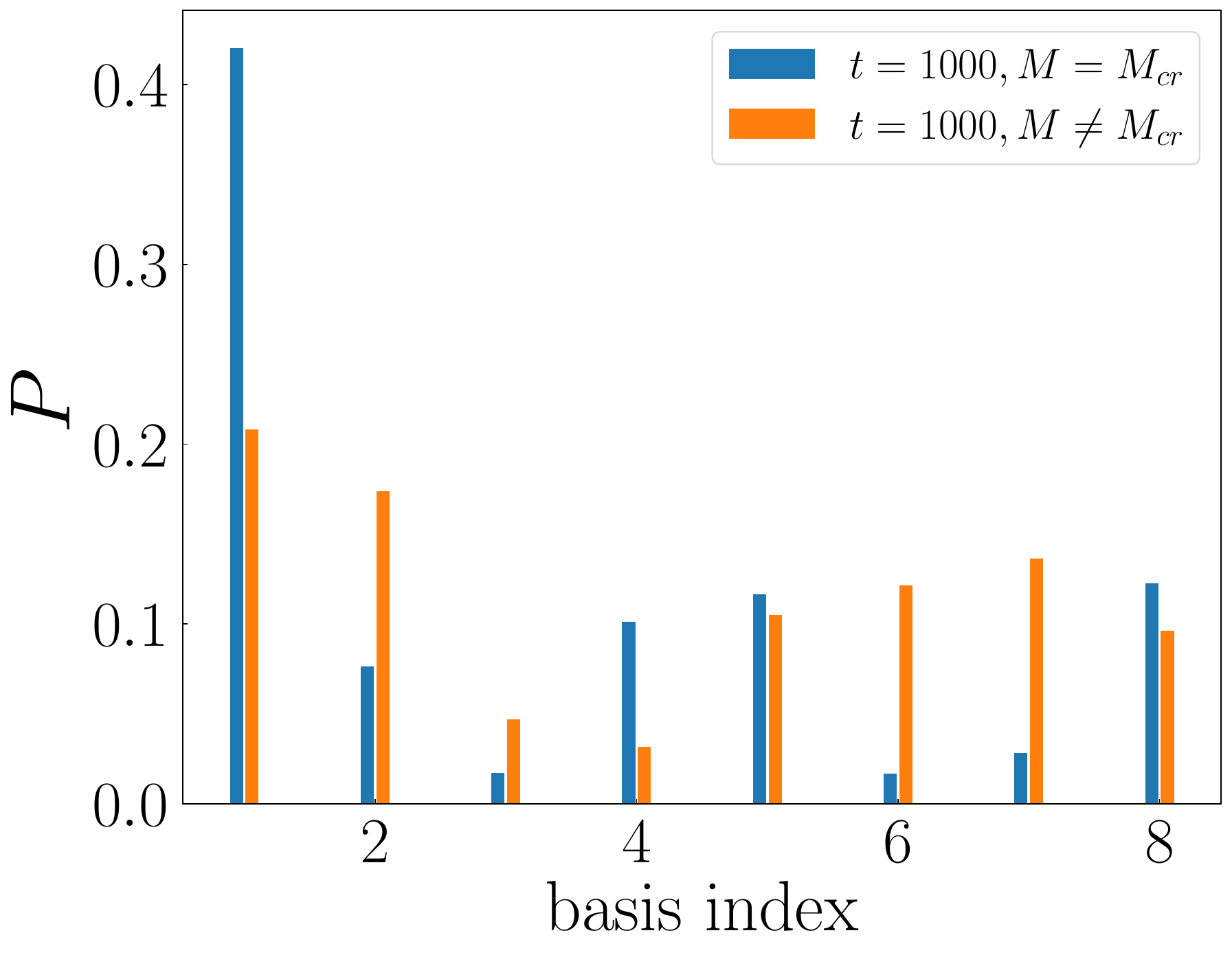}}\\
   \subfigure{%
   \includegraphics[scale=0.195]{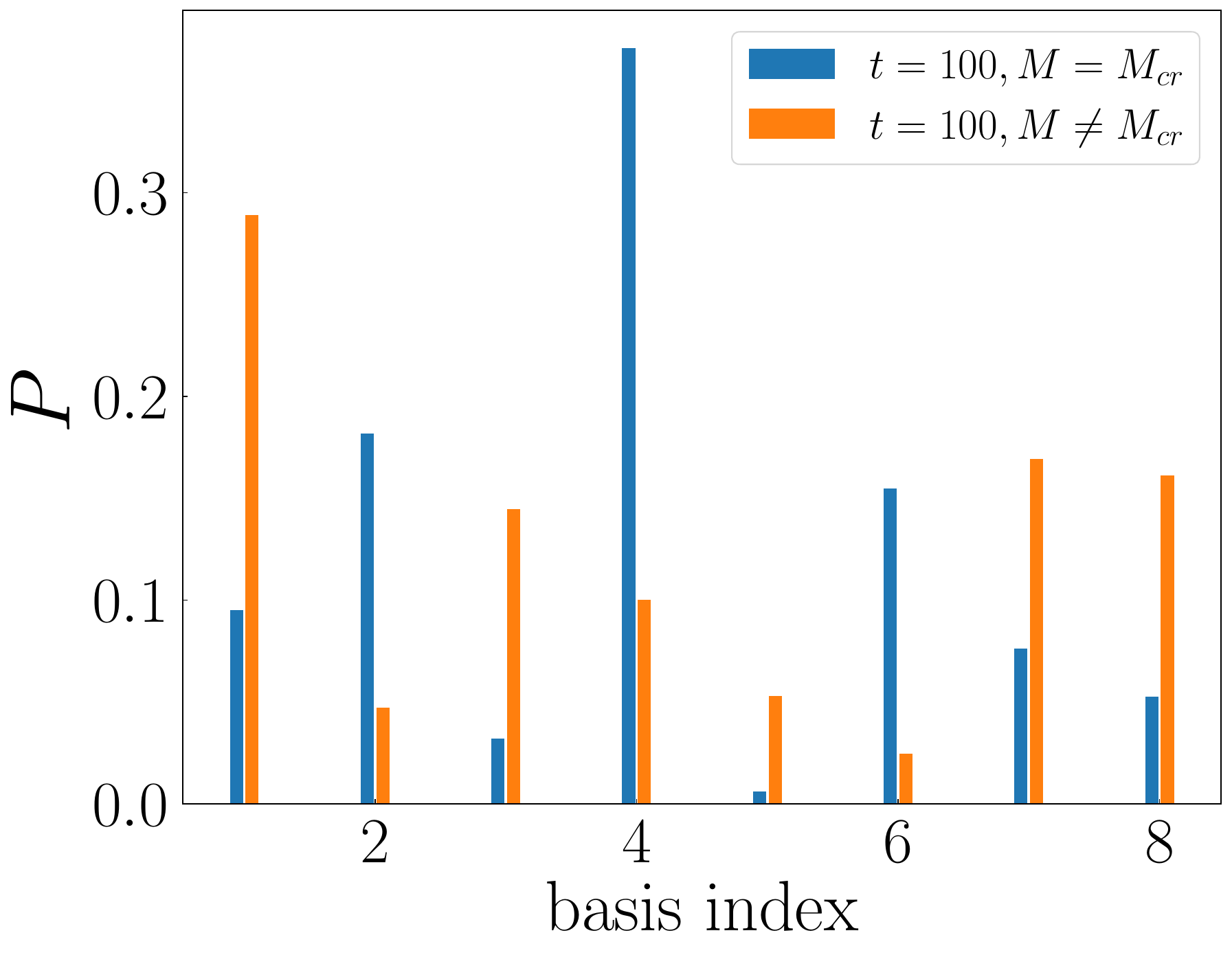}}
   \subfigure{%
   \includegraphics[scale=0.195]{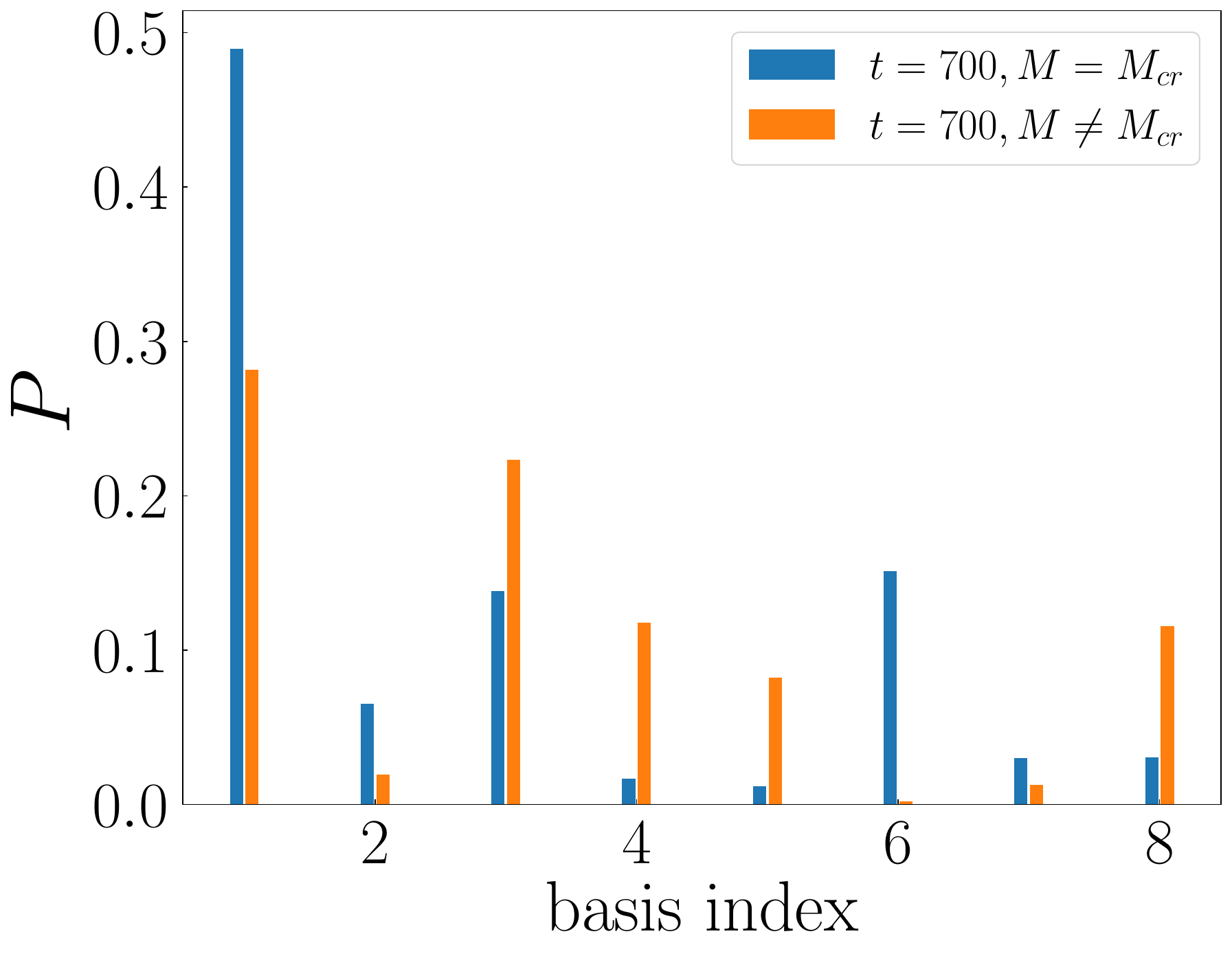}}
   \subfigure{%
   \includegraphics[scale=0.195]{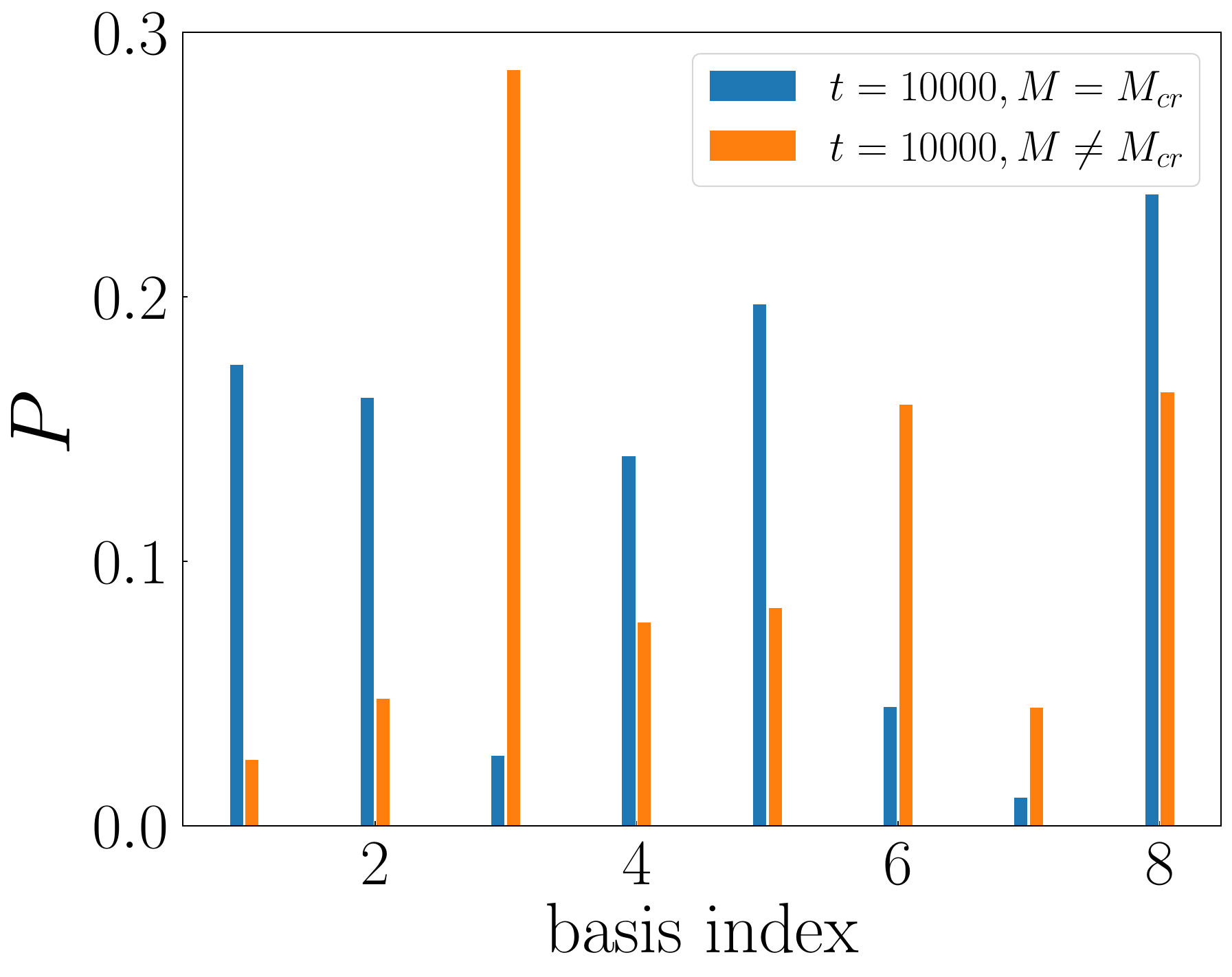}}
  \end{center}
      \caption{ Time dependence of excitation probability in the transformed Dicke basis (described as basis indices) at the crossing/resonance point $M= M_{cr}$ (in blue) and the non-resonant point (in orange).}
      \label{}
  \end{figure*}


\newpage

\begin{figure*}
  \begin{center}
  \subfigure[]{%
  \includegraphics[width=0.45\textwidth]{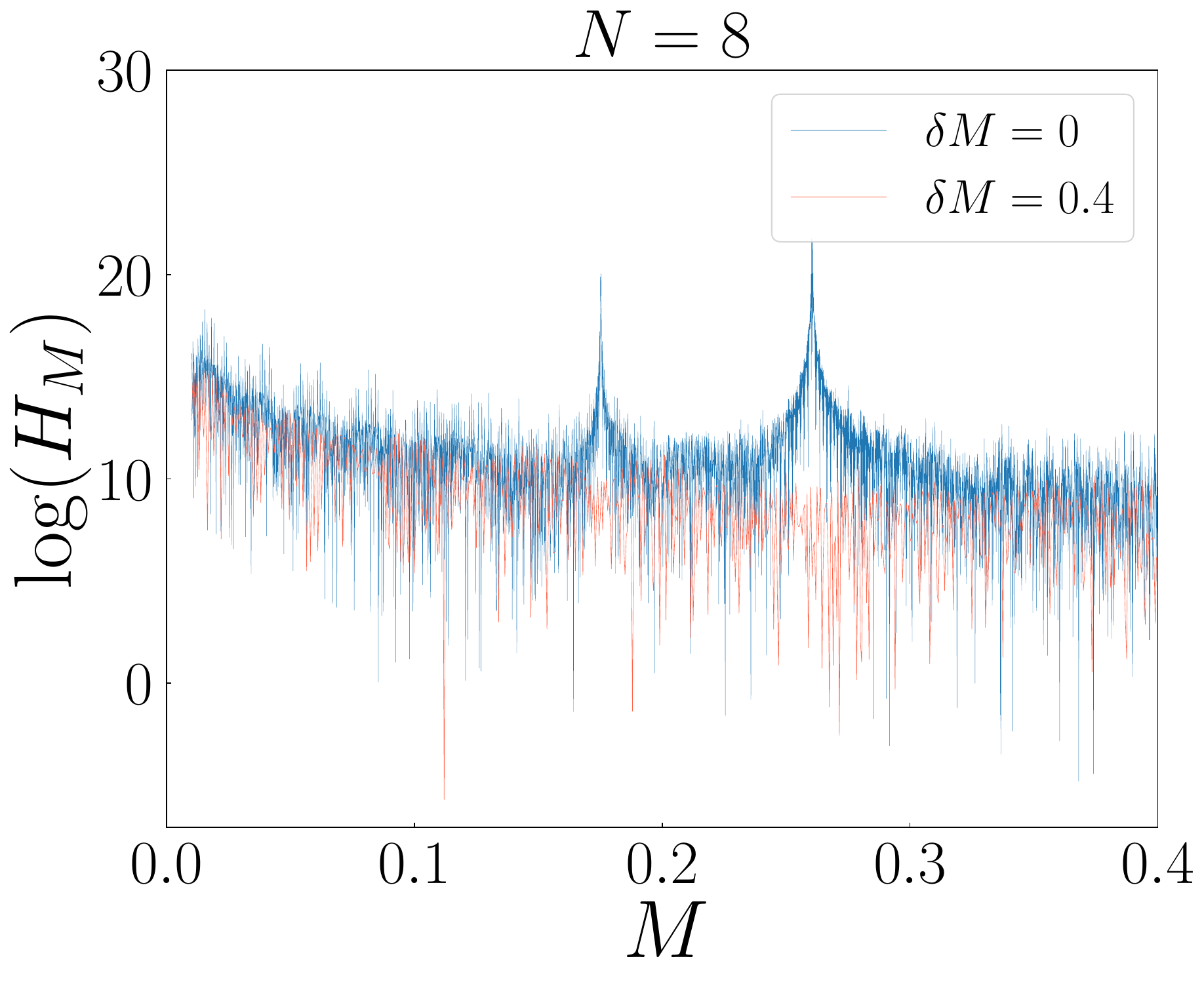}}
 \subfigure[]{%
  \includegraphics[width=0.44\textwidth]{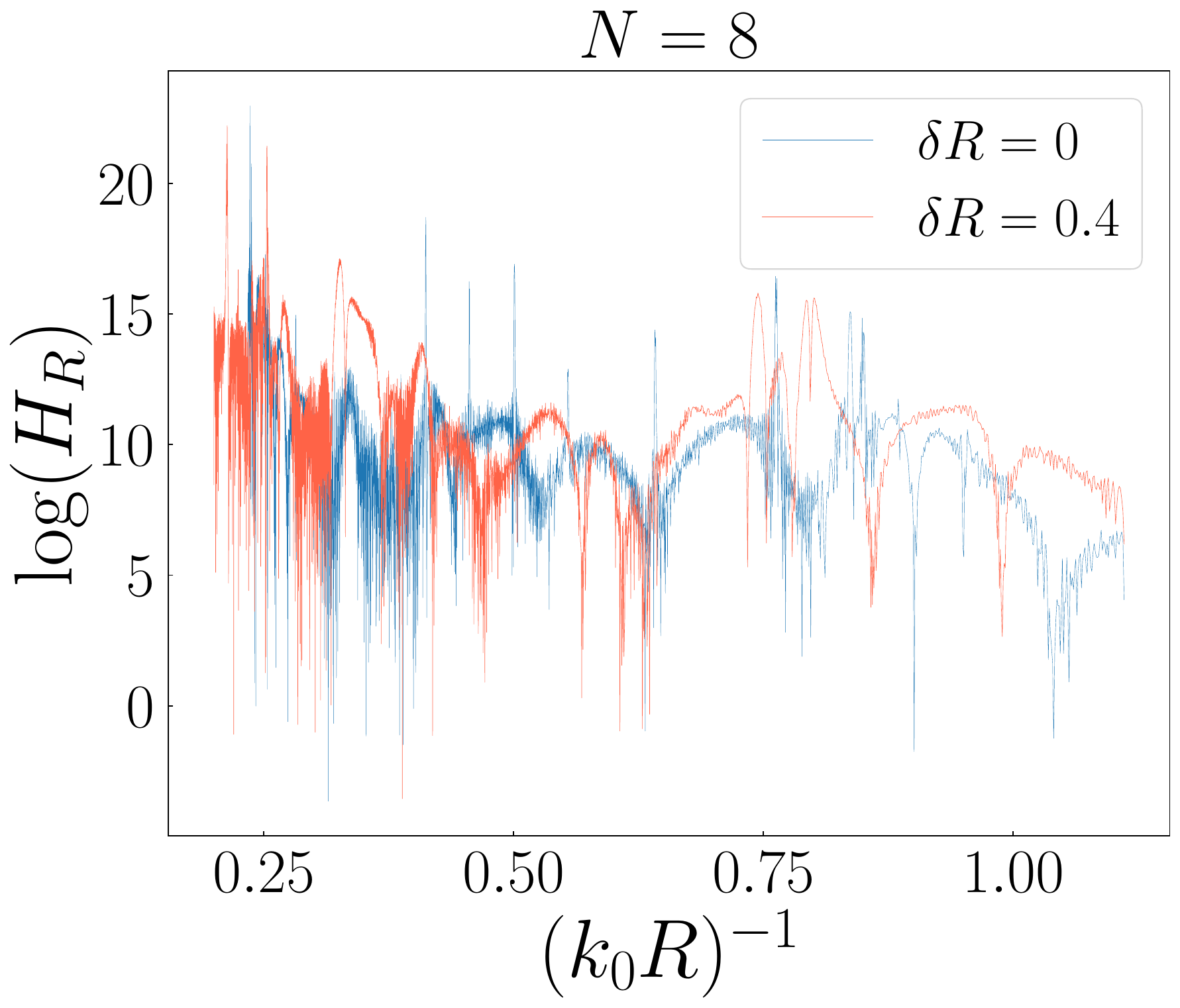}}
   \subfigure[]{%
  \includegraphics[width=0.48\textwidth]{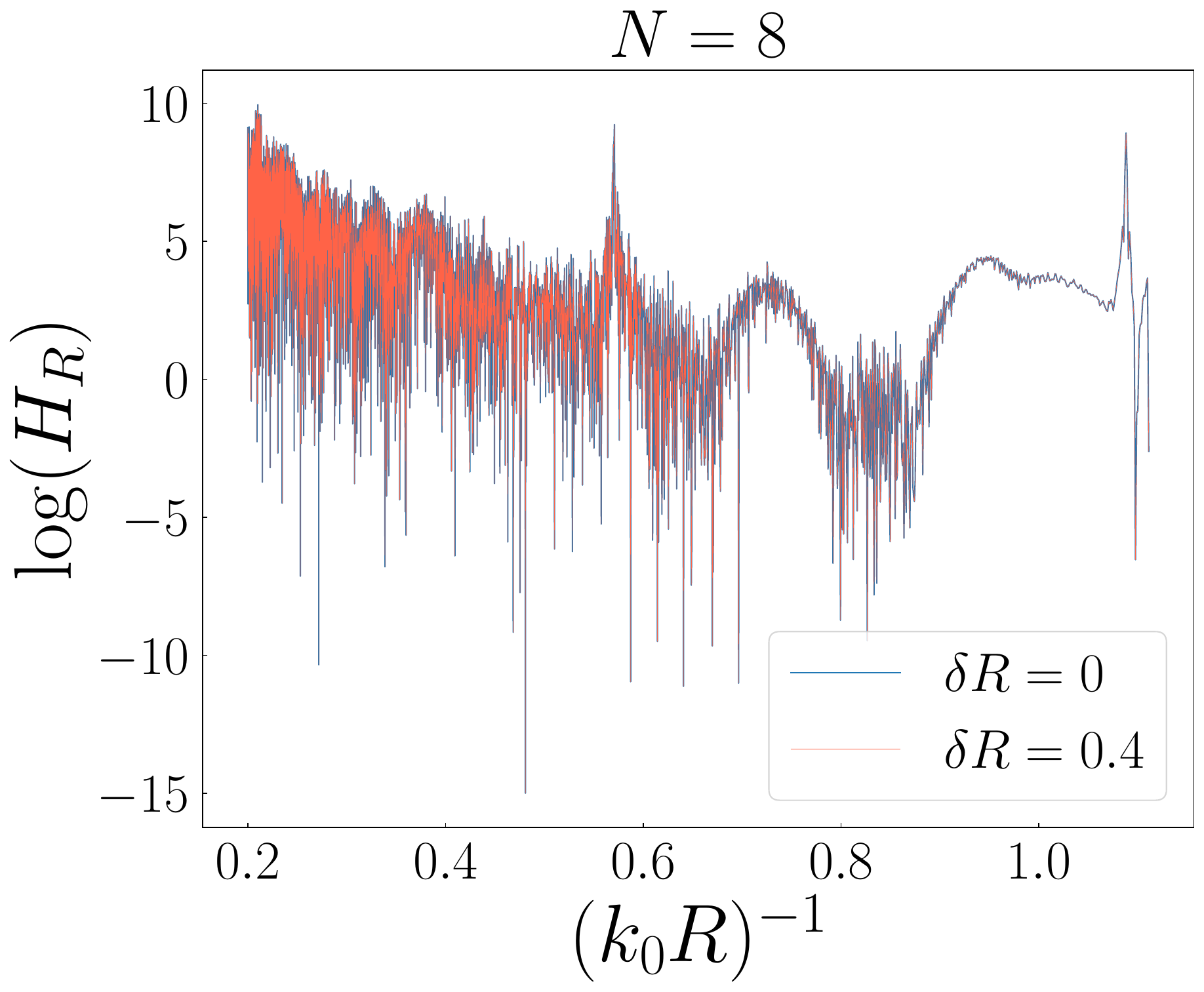}}
  \end{center}
      \caption{(a) The variation of the logarithm of the quantum Fisher information in the NN regime without dissipation for $N=8$ as a function of $M$ (dipole-dipole interaction strength \bblk{in units of $\gamma$}) in the presence of randomness (red solid line) and without randomness (black solid line). (b) Same for the long-range NNN interaction as a function of $1/(k_0 R)$ (for negligible dissipation $\gamma_{ij}/M \rightarrow 0$) in the presence of randomness (red solid line) and without randomness (black solid line). (c) Same for comparable dissipation ($\gamma_{ij}/M \rightarrow 1$) in the presence of randomness (red solid line) and without randomness (black solid line). The  Hamiltonian parameters $\omega - \omega_0 = 0.2 \gamma$,  $\kappa_j = 0.2 \gamma$,  $T =10000/\bblk{\gamma}$.}
      \label{fig:logfisher_N=8_WOR}
  \end{figure*}

\end{document}